\documentclass[twocolumn,a4paper,reprint,amssymb,aps,prd,nofootinbib,groupedaddress,superscriptaddress]{revtex4-1}
\pdfoutput=1
\usepackage{booktabs}
\usepackage{multirow}
\usepackage{hyperref} 
\usepackage{dcolumn}
\usepackage{bm}
\usepackage{amsmath}
\usepackage{mathtools}
\usepackage{amsfonts}
\usepackage{pbox}
\usepackage{color}
\usepackage{xcolor}
\usepackage{tabularx}
\usepackage{subfigure}
\usepackage{graphicx}
\usepackage{array}
\usepackage{psfrag}
\usepackage{lipsum}
\usepackage{enumitem}  
\usepackage{changes}
\usepackage{soul}

\def\D {\mbox{D}}

\begin{document}

\title{Simple-graduated dark energy and spatial curvature}

\author{Giovanni Acquaviva}
\email{gioacqua@utf.troja.mff.cuni.cz }
\affiliation{Institute of Theoretical Physics, Faculty of Mathematics and Physics, Charles University, CZ-180 00 Prague, Czech Republic}
\author{\"{O}zg\"{u}r Akarsu}
\email{akarsuo@itu.edu.tr}
\affiliation{Department of Physics, Istanbul Technical University, Maslak 34469 Istanbul, Turkey}
\author{Nihan Kat{\i}rc{\i}}
\email{nihan.katirci@itu.edu.tr}
\affiliation{Department of Physics, Istanbul Technical University, Maslak 34469 Istanbul, Turkey}
\author{J. Alberto Vazquez}
\email{javazquez@icf.unam.mx}
\affiliation{Instituto de Ciencias F\'isicas, Universidad Nacional Aut\'onoma de M\'exico, Cuernavaca, Morelos, 62210, Mexico}

\begin{abstract}
In this work, we first discuss the possibility that dark energy models with negative energy density values in the past can alleviate the $H_0$ tension, as well as the discrepancy with the baryon acoustic oscillation (BAO) Lyman-$\alpha$ data, both which prevail within the $\Lambda$CDM model. We then investigate whether two minimal extensions of the $\Lambda$CDM model, together or separately, can successfully realize such a scenario: (i) the spatial curvature, which, in the case of spatially closed universe, mimics a negative density source and (ii) simple-graduated dark energy (gDE), which promotes the null inertial mass density of the usual vacuum energy to an arbitrary constant---if negative, the corresponding energy density decreases with redshift similar to the phantom models, but unlike them crosses below zero at a certain redshift. We find that, when the Planck data are not included in the observational analysis, the models with simple-gDE predict interesting and some significant deviations from the $\Lambda$CDM model. In particular, a spatially closed universe along with a simple-gDE of positive inertial mass density, which work in contrast to each other, results in minor improvement to the $H_0$ tension. The joint dataset, including the Planck data, presents no evidence for a deviation from spatial flatness but almost the same evidence for a cosmological constant and the simple-gDE with an inertial mass density of order $\mathcal{O}(10^{-12})\,\rm eV^4$. The latter case predicts almost no deviation from the $\Lambda$CDM model up until today---so that it results in no improvement regarding the BAO Ly-$\alpha$ data--- except that it slightly aggravates the $H_0$ tension. We also study via dynamical analysis the history of the Universe in the models, as the simple-gDE results in futures different than the de Sitter future of the $\Lambda$CDM model.

\end{abstract}
%\date{\today}

\maketitle
\section{Introduction}
\label{sec:intro}

There is a growing consensus in recent years that the standard Lambda cold dark matter ($\Lambda$CDM) model must in fact be an approximation to a more realistic cosmological model that has not been understood yet \cite{DiValentino:2020vhf}. This new model is not expected to exhibit phenomenologically too drastic
%---but may be non-trivial---
deviations from the $\Lambda$CDM model, which is in excellent agreement with most of the currently available data \cite{Riess:1998cb,Ade:2015xua,Alam:2016hwk,Abbott:2017wau,Aghanim:2018eyx}, even if it could be conceptually very different. The recent developments, both theoretical (e.g., the de Sitter swampland conjecture \cite{Obied:2018sgi,Agrawal:2018own,Colgain:2018wgk,Heisenberg:2018yae,Akrami:2018ylq,Raveri:2018ddi,Cicoli:2018kdo,Colgain:2019joh}) and observational (e.g., the tensions prevailing within the $\Lambda$CDM model and preference for some unexpected and/or nontrivial extensions of this model; see Refs. \cite{Freedman:2017yms,Verde:2019ivm,DiValentino:2017gzb,Aubourg:2014yra,Zhao:2017cud,Bullock:2017xww,tension02,Mortsell:2018mfj,Dutta:2018vmq,Hee:2016ce,Vagnozzi:2019ezj,Handley:2019tkm,DiValentino:2019qzk,Akarsu:2019hmw,DiValentino:2020hov,Vagnozzi:2020zrh,Delubac:2014aqe,Dutta:2018vmq,Sahni:2014ooa,Visinelli:2019qqu,Poulin:2018zxs,Capozziello:2018jya,Wang:2018fng,Banihashemi:2018oxo,Banihashemi:2018has,Ye:2020btb,Perez:2020cwa,Bonilla:2020wbn,Vazquez:2020ani,Vagnozzi:2018jhn,DiValentino:2019exe,DiValentino:2019dzu,Calderon:2020hoc,Vazquez:2012ag,Paliathanasis:2020sfe,Akarsu:2020yqa,Efstathiou:2020wem,Vagnozzi:2020dfn,Banerjee:2020xcn} and \cite{DiValentino:2020zio,DiValentino:2020vvd,DiValentino:2020srs,Perivolaropoulos:2021jda} for more references), along with the notoriously challenging cosmological constant problem \cite{Weinberg:1988cp,Peebles:2002gy}, suggest that attaining this new model would not be a straightforward task. These tensions are of immense interest not only in cosmology but also in theoretical physics, as they could imply new physics beyond the well-established fundamental theories underpinning, or even extending, the $\Lambda$CDM model. The so-called $H_{0}$ tension---the deficit in the Hubble constant $H_0$ predicted by the Planck cosmic microwave background (CMB) data within the $\Lambda$CDM model \cite{Aghanim:2018eyx} when compared to the direct local distance ladder measurements \cite{Riess:2016jrr,Riess:2018byc,Riess:2019cxk,Freedman:2019jwv,Yuan:2019npk,Freedman:2020dne}--- among others is now described by many as a crisis. See Ref. \cite{DiValentino:2020zio} for a comprehensive list of references on the $H_0$ tension and Ref. \cite{DiValentino:2021izs} for a recent review on its possible solutions. It becomes quite perplexing as it worsens when the simplest minimally coupled single-field quintessence models are used instead of cosmological constant ($\Lambda$) and only partially improves when the simplest phantom (or quintom) models are used \cite{Vagnozzi:2018jhn,DiValentino:2019exe,DiValentino:2019dzu,Vazquez:2020ani,Banerjee:2020xcn}. Surprisingly, it has been reported that the $H_0$ tension---as well as a number of other low-redshift discrepancies---may be alleviated by a dynamical dark energy that assumes negative or vanishing energy density values at high redshifts \cite{Delubac:2014aqe,Aubourg:2014yra,Sahni:2014ooa,Mortsell:2018mfj,Poulin:2018zxs,Capozziello:2018jya,Wang:2018fng,Dutta:2018vmq,Banihashemi:2018oxo,Banihashemi:2018has,Visinelli:2019qqu,Akarsu:2019hmw,Ye:2020btb,Perez:2020cwa,Calderon:2020hoc,Paliathanasis:2020sfe,Bonilla:2020wbn,Vazquez:2012ag,Akarsu:2020yqa}. The fact that, when spatial curvature is allowed on top of the $\Lambda$CDM model, the Planck data combined with other astrophysical data [baryon acoustic oscillations (BAO), cosmic chronometers, etc.] favor spatial flatness ($\Omega_{k0}=0$) with extremely high precision, while the Planck data alone favor positive spatial curvature ($\Omega_{k0}<0$) might also be implying such dark energy models \cite{Aghanim:2018eyx,Handley:2019tkm,DiValentino:2019qzk,DiValentino:2020hov,Vagnozzi:2020zrh,Efstathiou:2020wem,Vagnozzi:2020dfn}.

%Moreover, the authors' bibliography can also be improved in the Introduction, when they mention that Planck 2018 prefers a spatially closed Universe [21-24]. I believe a more balanced discussion should also cite works which argue how breaking the geometrical degeneracy pushes the Planck results towards a spatially flat Universe. Examples of works which could be cited in this direction include Efstathiou & Gratton, MNRAS Letters 496 (2020) L91 [arXiv:2002.06892]; and Vagnozzi et al., ApJ 908 (2021) 84 [arXiv:2011.11645]

The constraint on $H_0$ from the CMB power spectrum is inferred from the distance between the acoustic peaks that measures the angular scale at last scattering $\theta_{\star} = r_{\star}/D_{\rm M}$, where $r_{\star}=\int^{\infty}_{z_{\star}}c_sH^{-1}{\rm d}z$ is the comoving sound horizon---determined by the pre-recombination physics---($c_s$ being the sound speed in the plasma) and $D_{\rm M}=\int_0^{z_{\star}}H^{-1}{\rm d}z$ is the comoving angular diameter distance---determined by the postrecombination physics---at last scattering, for which the redshift $z_{\star}\simeq 1090$ \cite{Dodelson03}. As dark energy [in general, described by an equation of state (EoS) parameter $w(z)\equiv p/\rho\sim-1$ with $\rho$ and $p$ being the energy density and pressure, respectively] is negligible at high redshifts, it is not expected to affect the prerecombination physics, viz., $r_{\star}$. Also, Planck satellite measures $\theta_{\star}$ very robustly (with an accuracy of 0.03\%) and almost independently of the cosmological model \cite{Aghanim:2018eyx}. Therefore, $D_{\rm M}$ must remain the same in different dark energy models (with some exceptions, e.g., early dark energy models \cite{Poulin:2018cxd}, which prescribe $D_{\rm M}\propto r_{\star}$). Thus, dark energy models that assume negative energy density values for $z>z_*$ ($z_*$ being the redshift at which the energy density crosses below zero) imply a reduction in $H(z)$ compared to the one in the $\Lambda$CDM model for $z>z_*$, more than the phantom/quintom dark energy models could achieve (for these models the energy density typically decreases and approaches zero with redshift). The compensation of this reduction to keep $D_{\rm M}$ unaltered results in an enhanced $H(z)$ for $z<z_*$, i.e., $H_0$ as well. Besides, this sign change in the dark energy density can even lead to a nonmonotonic behavior of $H(z)$ that fits better with the Lyman-$\alpha$ (Ly-$\alpha$) BAO measurements, e.g., those from
the Baryon Oscillation Spectroscopic Survey (BOSS) and from its extended version eBOSS in the Sloan Digital Sky Survey Data Release 14 (SDSS DR14) \cite{Agathe:2019vsu,Blomqvist:2019rah}, which (combined) present approximately $1.7\sigma$ tension with the prediction of the Planck (2015) \cite{Ade:2015xua} best-fit $\Lambda$CDM model. Such models were first suggested by the BOSS Collaboration when it was shown that the BAO peak position in the Ly-$\alpha$ at an effective redshift $z\sim2.34$ (BOSS DR11) \cite{Delubac:2014aqe} presents an approximately $2.5\sigma$ discrepancy with the CMB predictions from the Planck (2015) \cite{Ade:2015xua} best-fit $\Lambda$CDM model. Then, they reported a  dark energy density consistent with a positive cosmological constant for $z<1$, but with a preference for negative values for $z>1.6$, and argued that this discrepancy can be addressed by a nonmonotonic behavior of $H(z)$ at $z\sim2$ \cite{Aubourg:2014yra}. The Planck Collaboration (2018) \cite{Aghanim:2018eyx}  excludes the Ly-$\alpha$ BAO measurements from their default BAO compilation---as they do not significantly constrain the $\Lambda$CDM model, as well as its simple extensions, when CMB and galaxy BAO data are already used---and, mentioning from Ref.\cite{Delubac:2014aqe} that it is difficult to construct well-motivated extensions of the $\Lambda$CDM model that can resolve this tension, suggest further work is needed to assess whether it is a signature of new physics or not. Recently, the $\Lambda$ assumption was investigated by introducing the graduated dark energy (gDE) characterized by a minimal dynamical deviation from the null inertial mass density $\varrho=0$ (where $\varrho\equiv\rho+p$) of the $\Lambda$---or the usual vacuum energy of the quantum field theory (QFT)---in the form $\varrho\propto \rho^{\lambda}<0$ with $\lambda<1$ being a ratio of two odd integers, for which its energy density $\rho$ dynamically takes negative values in the past \cite{Akarsu:2019hmw}. For large negative values of $\lambda$, gDE creates a phenomenological model described by a smooth function that approximately describes the $\Lambda$ spontaneously flipping sign in the late Universe to become positive today. It was shown via the gDE that the joint observational data, including the Planck CMB and Ly-$\alpha$ BAO (BOSS DR11) data as well, suggest the $\Lambda$ spontaneously changed sign at redshift $z\approx 2.32$ and triggered the late-time acceleration, which alleviates both the $H_0$ tension and the discrepancy with the Ly-$\alpha$ BAO measurements. Currently, using the Ly-$\alpha$ BAO measurements in the final eBOSS (SDSS DR16), which contains all data from eBOSS and its predecessor, the BOSS, the tension with the prediction of the Planck (2015) \cite{Ade:2015xua} best-fit $\Lambda$CDM model is reduced to only approximately $1.5\sigma$ \cite{Alam:2020sor,duMasdesBourboux:2020pck}. On this account, the dark energy density that is almost constant today but assumes negative values in the past is not indispensable, yet it has potential to result in a better agreement with the existing observational data and the direct $H_0$ measurements.

Such scenarios elaborated by allowing negative energy densities in the Friedmann equation can be achieved as minimal extensions of the $\Lambda$CDM model. The simplest example of this type of contribution may be thought of as a single effective fluid defined as the total contributions of the positive cosmological constant and constant positive spatial curvature (which mimics a negative energy density source with $w=-1/3$) to the Friedmann equation. However, if its energy density  crosses below zero at $z\sim1.6$ in line with Ref. \cite{Aubourg:2014yra}, then the spatial curvature's density parameter today is required to be $\Omega_{k0}\sim-0.12$ (assuming $\Omega_{\rm m0}\sim0.3$ for dust), which contradicts the standard inflationary paradigm and the joint results of the recent Planck release \cite{Aghanim:2018eyx} suggesting spatial flatness to a $1\,\sigma$ accuracy of 0.2\%. Yet, recent observations hint toward a spatially closed universe, in particular, the Planck CMB power spectra prefer a closed universe at more than 99\% confidence level (assuming the nominal likelihood) \cite{DiValentino:2019qzk,DiValentino:2020hov,Handley:2019tkm,DiValentino:2020hov,Vagnozzi:2020zrh}. If we adhere to the inflationary paradigm, with a faithful assumption on the spatial flatness, such behavior can be achieved in a minimalist way by promoting phenomenologically the null inertial mass density $\varrho=0$ of the usual vacuum energy of the QFT to a negative constant $\varrho=\rm{const}<0$---henceforth, \textit{simple-graduated dark energy} (simple-gDE) corresponding to the case $\lambda=0$ of the gDE. Its energy density decreases with increasing redshift like phantom dark energy models, but unlike them crosses below zero at a certain redshift. The source satisfying $\varrho=\rm{const}$ has recently been of interest to many as it can resemble $\Lambda$ today, while leading to a future singularity dubbed as the little sibling of the big rip (LSBR) for $\varrho=\rm{const}<0$ or a finite future bounce for $\varrho=\rm{const}>0$ \cite{Bouhmadi-Lopez:2014cca,Albarran:2016mdu,Bouali:2019whr}. Such a contribution to the Friedmann equation may also be obtained from a modified gravity; see, for instance Ref. \cite{Akarsu:2019ygx}, where it arises from barotropic perfect fluid via the energy-momentum squared gravity of the form $f(T_{\mu\nu}T^{\mu\nu})\propto \ln(\lambda\,T_{\mu\nu}T^{\mu\nu})$. However, if its energy density is positive today and crosses below zero at $z\sim1.6$, then its EoS parameter today yields $w_0\sim -1.35$, which obviously is not expected to be permitted by the observational data. On the other hand, in case $\Omega_{k0}$ and $\varrho=\rm const$ are simultaneously allowed to be free parameters on top of the $\Lambda$CDM model, it can be possible to draw both $\Omega_{k0}$ and $w_0$ parameters to observationally reasonable values, though they together can still lead to sufficiently large deviations from the $\Lambda$CDM model resulting in better agreement with the observational data. Yet, at the end of the day, we may find ourselves with a completely different outcome. In particular, given that there exist model-independent inferences of $H_0$ from the inverse distance ladder, which show that combined dataset of BAO and Type Ia Supernovae (SN) with/without Cepheids prefer low $H_0$ values independently of Planck data and the adopted dark energy model suggesting late-time modifications alone is unable to alleviate the $H_0$ tension and turned attention to early-time modifications (such as early dark energy) which lowers the sound horizon \cite{Bernal:2016gxb,Lemos:2018smw,Aylor:2018drw,Knox:2019rjx,Efstathiou:2021ocp}. Also, notice that our discussions above also imply that the spatial curvature with $\Omega_{k0}>0$ (spatially open universe) and the simple-gDE with $\varrho>0$ (in this case the energy density increases with increasing redshift, likewise the quintessence dark energy models) would exacerbate both the $H_0$ tension and discrepancy with the BAO Ly-$\alpha$ measurements, both which prevail within the $\Lambda$CDM model. In what follows, we explore in detail whether the spatial curvature ($\Omega_{k0}$) and the simple-gDE ($\varrho=\rm const$) extensions, separately or simultaneously, of the standard $\Lambda$CDM model result in improvements in fitting the observational data. We also discuss the implications of our observational constraints on the past and future history of the Universe and nature of the vacuum energy. 

\section{Model}
\label{sec:DEmodel} 
The energy-momentum tensor describing an isotropic perfect fluid can be decomposed relative to a unique $4$-velocity $u^\mu$ (satisfying $u_\mu u^{\mu}=-1$ and $\nabla_{\nu}u^{\mu}u_{\mu}=0$) in the form of $T_{\mu\nu}=(\rho+p) u_\mu u_\nu+p g_{\mu\nu}$, where $\rho$ is the (relativistic) energy density relative to $u^\mu$, $p$ is the isotropic pressure and $g_{\mu\nu}$ is the metric tensor. In general relativity (GR)---described by the Einstein field equations $G_{\mu\nu} =-T_{\mu\nu}$---the set of equations arises from the twice-contracted Bianchi identity implying $\nabla_{\mu}G^{\mu\nu}=0$ and hence resulting in $\nabla_{\mu}T^{\mu\nu} = 0$. Projecting parallel and orthogonal to $u^{\mu}$, we obtain the energy and momentum conservation equations, correspondingly,
\begin{eqnarray}
\label{eqn:constraint}
\dot\rho+\Theta\varrho=0\quad\textnormal{and}\quad
{\rm D}^\mu p+\varrho\dot{u}^\mu=0,
\end{eqnarray}
  where $\Theta={\rm D}^\mu u_\mu$ is the volume expansion rate, a dot denotes the derivative with respect to the comoving time $t$, and we have used $\nabla_{\nu}u_{\mu}=\D_{\nu}u_{\mu}-\dot{u}_{\mu} u_{\nu}$  \cite{EllisRC,Ellis:1998ct}. Notice, in the momentum conservation equation in \eqref{eqn:constraint}, that ${\rm D}^\mu p$ is the pressure gradient and $\dot{u}^\mu$ is the $4$-acceleration, and therefore $\varrho=\rho+p$ defines \textit{the inertial mass density}.

The usual vacuum energy of the QFT (described by the  EoS $p_{\rm cc}=-\rho_{\rm cc}$) corresponds to the source that yields null-inertial mass density,
\begin{equation}
\label{eqn:uvac}
    \varrho_{\rm cc}=0,
\end{equation}
for which $\rho_{\rm cc}={\rm const}>0$, namely, the energy density is a constant---via the energy conservation equation in \eqref{eqn:constraint}---and supposed to be positive as suggested by the cosmological observations.
%\jav{question: if $\varrho$ is different from zero, then D$^\mu p \ne 0$. Is there any implication in that?, what happens with $p$?}
The simplest phenomenological generalization of the usual vacuum energy \eqref{eqn:uvac} is then to promote its \textit{null inertial mass density} to an arbitrary constant,
\begin{equation}
\label{eqn:GDE}
\varrho_{\rm ci}=\rm const,
\end{equation}
for which the energy density $\rho_{\rm ci}$ (supposed to be positive today, i.e., $\rho_{\rm ci0}>0$) and the pressure $p_{\rm ci}$ are not necessarily constant---here and in what follows the subscript $0$ attached to any quantity denotes its present-day ($z=0$) value. It is worth noting that this promotion corresponds to taking the inertial mass density, instead of vacuum energy density (or $\Lambda$), as one of the constants of nature. We do not consider the possibility of $\rho_{\rm ci0}<0$ throughout our study, as it obviously contradicts the observations. The energy density of this source, which we call simple-graduated dark energy, reads
\begin{align}
\label{eqn:rhogde}
\rho_{\rm ci}=\rho_{\rm ci0}+3\varrho_{\rm ci} \ln (1+z),
\end{align}
which satisfies the EoS parameter ($w\equiv-1+\varrho/\rho$)
\begin{equation}
\label{eq:eos}
  w_{\rm ci}=-1+\frac{1+w_{\rm ci0}}{1+3\left(1+w_{\rm ci0}\right)\ln(1+z)},
\end{equation}
where $z=-1+\frac{a_0}{a}$ is the redshift with $a$ being the scale factor of the Robertson-Walker (RW) metric. This source, regardless of the sign of $\varrho_{\rm ci}$, eventually becomes indistinguishable from the $\Lambda$ in the past (say, $w_{\rm ci}\approx-1$ for $z\gg0$), and thus the extension of the $\Lambda$CDM model via this source approximates indefinitely close to the $\Lambda$CDM model as the dust dominates it in the past. Yet, as the future Universe will eventually be dominated by this source, the future will be drastically different depending on the sign of $\varrho_{\rm ci}$; the Universe hits a bounce ($H=0$) in the finite future if $\varrho_{\rm ci}>0$ and exhibits LSBR singularity in the infinite future if $\varrho_{\rm ci}<0$ \cite{Bouhmadi-Lopez:2014cca}. The latter case, $\varrho_{\rm ci}<0$, is of particular interest to us, as in this case $\rho_{\rm ci}$ decreases as $z$ increases; namely, the source exhibits a phantomlike behavior as the logarithmic term (the new term that arises due to the deviations from null inertial mass density) dynamically screens $\rho_{\rm ci0}$ in the finite past ($z>0$). However, in contrast to the usual phantom dark energy models (described by $w<-1$ with $\rho>0$), (i) its energy density does not asymptotically approach zero as $z$ increases but crosses below zero at
\begin{equation}
\label{eq:zpole}
z_{\rm ci*}=-1+{\rm e}^{-\frac{1}{3(1+w_{\rm ci0})}}
\end{equation}
and then keeps growing in negative values, and (ii) its EoS parameter yields $w_{\rm ci0}<-1$ for $z<z_{\rm ci*}$ and $w_{\rm ci0}>-1$ for $z>z_{\rm ci*}$. Unless $w_{\rm ci0}\neq -1$, it yields $w_{\rm ci}\rightarrow-1$ both in the far future ($z\rightarrow-1$) and in the very early Universe ($z\rightarrow\infty$) and exhibits a pole at $z_{\rm ci*}$, i.e., when the energy density crosses zero, which is in the finite past for $w_{\rm ci0}<-1$ and in the finite future for $w_{\rm ci0}>-1$. The case $w_{\rm ci0}=-1$ corresponds to the usual vacuum energy---for this, we obtain either $z_*=-1$ or $z_*=\infty$; both transitions imply such a thing would never happen. One may have noticed that simple-gDE \eqref{eqn:rhogde} with $\varrho_{\rm ci}<0$ is reminiscent of the phenomenological emergent dark energy (PEDE) \cite{Li:2019yem} described by the energy density $\rho=\rho_0-\rho_0\tanh{\left(\log_{10}(1+z)\right)}$ that decreases monotonically with increasing redshift due to the $\tanh$ term. However, in contrast to the simple-gDE, PEDE does not introduce an extra free parameter compared to the $\Lambda$CDM model and, like in the usual phantom dark energy models, yields positive definite energy density (vanishing for arbitrarily large redshifts) controlled by a particular EoS parameter that is $-1.14$ today and increases from $-1.29$ in the past to $-1$ in the future. One may further see Refs. \cite{Li:2020ybr,Yang:2021eud} for one-parameter extension of PEDE and compare with simple-gDE.

The spatial curvature of the RW metric can effectively be treated as a source described by an EoS parameter $w_k=-1/3$ and the corresponding energy density reads
\begin{equation}
\label{eqn:ksource}
    \rho_{k}=\rho_{k0}(1+z)^2,
\end{equation}
for which $\rho_{k0}>0$, $\rho_{k0}=0$, and $\rho_{k0}<0$ correspond to spatially open, flat, and closed universes, respectively. 
Now, we define the effective source ($k{\rm ci}$) made up of the constant inertial mass density source ($\rm ci$) and the spatial curvature ($k$), for which the energy density, $\rho_{k\rm ci}\equiv \rho_{\rm ci}+\rho_{k}$, reads
\begin{align}
\label{eq:rhotilde}
\rho_{k\rm ci}=\rho_{\rm ci0}\left[1+3(1+w_{\rm ci0})\ln (1+z)\right]+\rho_{k0}(1+z)^{2},
\end{align}
from \eqref{eqn:rhogde} and \eqref{eqn:ksource}. 
Note that, in the past ($z>0$), it never scales faster than either the dust energy density $\rho_{\rm m}\propto(1+z)^3$ or the radiation energy density $\rho_{\rm r}\propto(1+z)^4$, so that we can always recover the standard cosmology at sufficiently large redshifts and thus leave the usual prerecombination physics and dynamics of the Universe---as well as the comoving sound  horizon at last scattering $r_{\star}$---unaltered. It crosses zero at the redshift 
\begin{align}
z_{k{\rm ci}*} =\left\{
  \begin{array}{@{}ll@{}}
   \begin{rcases} -1+{\rm e}^{-\frac{1}{3(1+w_{\rm ci0})}-\frac{1}{2} W_0(x)}\,\,\,\, \end{rcases} \quad \textnormal{for} \quad w_{\rm ci0}<-1,\\
     \begin{rcases}-1+\sqrt{-\frac{\Omega_{{\rm cc}0}}{\Omega_{k0}}}   \,\,\,\,\, \quad \quad\quad\quad\,\end{rcases} \quad \textnormal{for} \quad w_{\rm ci0}=-1,\\
      \begin{rcases}
     -1+{\rm e}^{-\frac{1}{3(1+w_{\rm ci0})}-\frac{1}{2}  W_{-1}(x)}\\
     -1+{\rm e}^{-\frac{1}{3(1+w_{\rm ci0})}-\frac{1}{2} W_0(x)}
     \end{rcases} \quad \textnormal{for} \quad w_{\rm ci0}>-1,
  \end{array}\right. 
  \label{eq:tildezpole}
\end{align}
where $W_{0}(x)$ and $W_{-1}(x)$ are the two real branches of the Lambert $W(x)$ function with the argument defined as $x=\frac{\Omega_{k0}}{\Omega_{\rm ci0}} \frac{2}{3(1+w_{\rm ci0})} {\rm e}^{-\frac{2}{3(1+w_{\rm ci0})}}$. $\Omega_{i}=\rho_{i}/\rho_{\rm c}$ is the density parameter of the $i{\rm th}$ fluid, and $\Omega_{k}=\rho_{k}/\rho_{\rm c}$ corresponds to the spatial curvature, with $\rho_{\rm c}=3H^2$ being the critical energy density. 

\begin{figure}[t!]
  \includegraphics[width=0.4\textwidth]{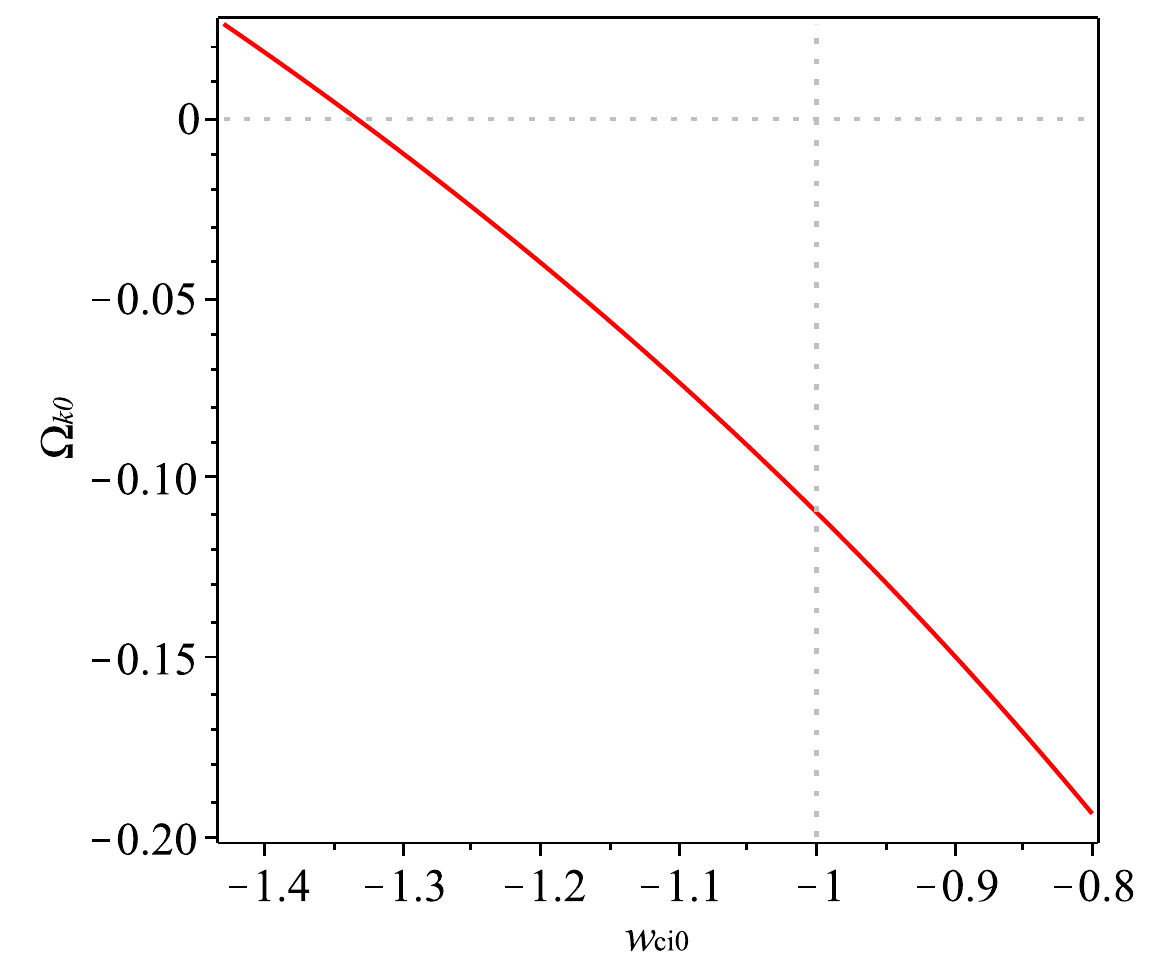}
  \caption{$\Omega_{k0}$ vs  $w_{\rm ci0}$ satisfying $z_{k{\rm ci}*}=1.72$ along with $\Omega_{\rm m0}=0.30$. The point $\{\Omega_{k0},w_{\rm ci0}\}=\{0,-1\}$ corresponds to the $\Lambda$CDM model.}
  \label{fig:range}
\end{figure}

We calculate from \eqref{eq:rhotilde} that $\rho_{k\rm ci}$ crossing below zero, for example, at $z_{k{\rm ci}*}={\rm e}-1\approx1.72$ (along with $\Omega_{\rm m0}=0.30$) in line with Ref. \cite{Aubourg:2014yra} requires $\Omega_{k0}=0.7+5.17/(3 w_{\rm ci0}-3.39)$. This implies significantly large deviations from the spatial flatness ($\Omega_{k0}=0$) or the cosmological constant ($w_{\rm ci0}=-1$), when only one of these is allowed to be a free parameter. Namely, we have $\Omega_{k0}=-0.11$ for $w_{\rm ci0}=-1$ and $w_{\rm ci0}=-1.33$ for $\Omega_{k0}=0$. Such large deviations are not expected to be permitted by the observational data. However, as may be seen in Fig.~\ref{fig:range} in case $\Omega_{k0}$ and $w_{\rm ci0}$ are simultaneously allowed to be free parameters, it is possible to bring each of them to more reasonable values; namely, we have $\Omega_{k0}=-0.07$ along with $w_{\rm ci0}=-1.1$. Thus, it is possible to have an overall large deviation from the standard $\Lambda$CDM model, while keeping the deviations from spatial flatness and cosmological constant at relatively moderate levels.

The Friedmann equation giving the complete description of the model under consideration here reads
\begin{align}
\label{eq:fried}
\frac{H^2}{H_0^2}=\,&\Omega_{\rm ci0}\left[1+3(1+w_{\rm ci0}) \ln (1+z) \right]+\Omega_{k0}(1+z)^2 \nonumber \\
&+\Omega_{\rm m0}(1+z)^3+\Omega_{\rm r0}(1+z)^4,
\end{align}
where the density parameters, except the one corresponding to the spatial curvature $\Omega_{k0}$, are positive definite. The $H=H(z)$ here can even exhibit a nonmonotonic behaviour, which may then reconcile the model with the Ly-$\alpha$ BAO measurements, as discussed in the Introduction. One may check that \eqref{eq:fried} has a minimum at $z=z_{\rm min}$ satisfying the relation (neglecting radiation as $\Omega_{\rm r 0}\sim10^{-4}$ for $z\sim1$)
\begin{align}
3\Omega_{\rm m0}(1+z_{\rm min})^3+2\Omega_{k0}(1+z_{\rm min})^2=-3\Omega_{\rm ci0}(w_{\rm ci0}+1).
\end{align}
And, it passes through its minimum in the past, i.e.,
\begin{align}
\label{eq:wcond}
z_{\rm min}>0, \quad \textnormal{if} \quad w_{\rm ci0}<\frac{3-\Omega_{k0}}{3(\Omega_{k0}+\Omega_{\rm m0}-1)}.
\end{align}
It implies $w_{\rm ci0}<-1.43$ for the spatially flat universe ($\Omega_{k0}=0$), while, e.g., $w_{\rm ci0}<-1.28$ for the spatially closed universe with $\Omega_{k0}=-0.11$. For example, $z_{\rm min}=1$ requires $w_{\rm ci0}=-4.43$ for $\Omega_{k0}=0$, and $w_{\rm ci0}=-3.60$ for $\Omega_{k0}=-0.11$. Such large deviations from the cosmological constant and/or the spatial flatness, may be implying that, when confronted with the observational data, the effective source that can assume negative density values in the finite past, i.e., $\rho_{k{\rm ci}}$ given in \eqref{eq:rhotilde}, would not be able to lead to a nonmonotonic evolution of $H(z)$ at about $z\sim1$ so as to reconcile it with the Ly-$\alpha$ BAO measurements. Of course, this can still result in a partial improvement in fitting the Ly-$\alpha$ BAO measurements, while predicting larger $H_0$ values (with respect to the $\Lambda$CDM model) toward addressing the so-called $H_0$ tension. Yet, a conclusive answer as to whether the model under consideration here \eqref{eq:fried} does better than the $\Lambda$CDM model cannot be given unless we rigorously confront the model with the latest observational data.

\section{Observational constraints}
\begin{table*}[ht!]
  \caption{Constraints ($68\%$ CL) on the parameters using the combined BAO+SN+$H$ and BAO+SN+$H$+PLK datasets. Before the two last rows, $-2\ln{\mathcal{L}_{\rm max}}$ is used to compare best fit with respect to the standard $\Lambda$CDM model. The last rows contain the Bayesian evidence $\ln \mathcal{Z}$ and the relative Bayesian evidence with respect to the standard $\Lambda$CDM model $\Delta\ln \mathcal{Z}=\ln \mathcal{Z}-\ln \mathcal{Z}_{{\Lambda}\rm CDM}$.
  %$\rho_{\rm ci 0}=1.879\, \Omega_{\rm ci0 }h_0^2\times10^{-29}\, \rm g\, cm^{-3}$ and  $\varrho_{\rm ci}=\rho_{\rm ci0}+p_{\rm ci0}=\rho_{\rm ci0}(1+w_{\rm ci0})$. \jav{done} $\rho_{{\rm cr}0}\sim 0.869\times10^{-29}\, \rm g\, cm^{-3}$
  }
  \label{tab:priors}
	\scalebox{0.74}{
	\setlength\extrarowheight{1.5pt}
  \begin{tabular}{lcccccccc}
  	\hline
    \toprule
    \multicolumn{1}{l}{Dataset} & \multicolumn{4}{c}{\textbf{BAO+SN+$H$}} & \multicolumn{4}{c}{\textbf{BAO+SN+$H$+PLK}}   \\  \hline
      & \textbf{{$\Lambda$CDM}} &  $\quad$ \textbf{$o\Lambda$CDM} & $\quad$\textbf{DE} & $\quad$ \textbf{$o${\rm DE}} & $\quad$\textbf{{$\Lambda$CDM}} &$\quad$ \textbf{$o\Lambda$CDM} &$\quad$ \textbf{DE} & $\quad$ \textbf{$o${\textbf{\rm DE}}}  \\ 	\hline \hline
      \midrule
      \vspace{0.1cm}
$\Omega_{\rm m0}$ & $0.307\pm0.014$ &  $0.310 \pm 0.020$ & $0.304 \pm 0.015$ &   $0.322 \pm 0.022$ 
    & $0.3005 \pm 0.0068$    & $0.3009 \pm 0.0067$ & $ 0.3070 \pm 0.0088$  & $0.3071 \pm 0.0091$  \\ \vspace{0.1cm}

$\Omega_{\rm b0}h_0^2$ &$0.02204 \pm 0.00047$& $0.02204 \pm 0.00046$ & $0.02204 \pm 0.00047$ & $0.02204\pm 0.00045$ 
 & $0.02245 \pm 0.00015$ & $0.02237 \pm 0.00017$& $0.02242 \pm 0.00015$ & $ 0.02241\pm 0.00017$   \\ \vspace{0.1cm}

$h_0$  & $0.6827 \pm 0.0088$    &$0.6862 \pm 0.0268$  & $0.6706 \pm 0.0202$ & $0.6884 \pm 0.0260$   
&    $0.6829 \pm 0.0052$& $ 0.6849 \pm 0.0067$ & $0.6772 \pm 0.0097$ & $0.6773 \pm 0.0099$  \\ \vspace{0.1cm}

 $w_{\rm ci0}$  &     $-1$  & $-1$ &$ -0.937 \pm 0.084$ &$-0.872 \pm 0.097$ 
&$-1$  &  $-1$  &$-0.948 \pm 0.041$ & $-0.951\pm 0.045$    \\ \vspace{0.1cm}

$\Omega_{k0}$  & ---   & $-0.011 \pm 0.077$  & --- & $-0.122 \pm 0.117$  
&     --- & $0.0012\pm 0.0018$ & --- & $-0.0001 \pm 0.0019$    \\ 
\hline
\vspace{0.1cm}
 $\varrho_{\rm ci}\times 10^{31}$ $[\rm g\, cm^{-3}]$  &     $0$  & $0$ &$ 3.46\pm4.76$ &$7.65\pm 5.72$ 
&$0$  &  $0$  &$3.06\pm2.28$ & $2.85\pm2.58 $    \\ \vspace{0.1cm}

$\Omega_{\rm ci0}$ &   $0.693\pm0.014$ & $0.700\pm 0.064$  & $0.696\pm 0.015$ & $0.800\pm 0.101$ 
& $0.6994\pm 0.0068$ & $0.6977 \pm 0.0065$  &$0.6929\pm 0.0088$ & $0.6929\pm 0.0095$   \\  \vspace{0.10cm} 

 $\Omega_{k{\rm ci}0}$ &   ---   & $0.690 \pm 0.020$  & --- & $0.678\pm 0.022$ & ---   & $0.6991 \pm 0.0067$  & --- & $0.6928\pm 0.0091$ \\

 $z_{{\rm ci}*}$ &    ---  & ---  & $<-0.96$ or $\gtrsim10^7$  & $<-0.78$ & ---  & ---  & $<-0.99$ & $<-0.99$ \\

 $z_{k{\rm ci}*}$ ($z_{k{\rm cc}*}$) &  ---  & $>1.26$ & --- & $>0.92$ & ---  & $>9.62$   & --- & $>6.64$  \\

%$z_{k{\rm ci}*}$ ($z_{k{\rm cc}*}$) &  ---  & $>8.81$ & --- & $2.07\pm1.51$ & ---  & $>1154.52$   & --- &  $>18.32$  \\
  
%$z_{k{\rm ci}*}$ ($z_{k{\rm cc}*}$) &  ---  & $3.83\pm 3.08$ & --- & $2.45\pm 2.77$ & ---  & $>16.10$   & --- &  $>371.16$  \\
 
  \hline
 \vspace{0.10cm} 
   $-2\ln{\mathcal{L}_{\rm max}}$ & $58.97$ & $58.96$ & $58.28$ & $56.91$ & $60.46$ & $59.27$ & $58.24$ & $58.24$  \\ \vspace{0.10cm}
  $\ln \mathcal{Z}$ & $-36.54\pm0.19$         & $-38.38\pm0.21$    & $-37.96\pm0.21$ &  $-38.00\pm0.21$      &     $-42.02\pm0.26$ & $-43.78\pm0.26$    & $-42.19\pm0.25$ &  $-44.13\pm0.27$ \\
  
 $\Delta\ln \mathcal{Z}$ & $0$         & $-1.84\pm0.28$    & $-1.42\pm0.28$ &  $-1.46\pm0.28$      &     $0$ & $-1.76\pm0.37$    & $-0.17\pm0.36$ &  $-2.11\pm0.37$ \\ 

    \bottomrule
    \hline 
    \hline
  \end{tabular}}
\end{table*}

\label{sec:observations}

We perform a parameter estimation and provide observational constraints from the latest data on the free parameters of the models under consideration---in the rest of the paper, $o\Lambda$CDM refers to the model including the spatial curvature on top of the standard $\Lambda$CDM model and similarly $o{\rm DE}$ and ${\rm DE}$ refer to the models considering the simple-gDE (characterized by constant inertial mass density) with and without the spatial curvature, respectively. To explore the parameter space, we make use of the new version of the SimpleMC code \cite{Anze}, initially released in Ref. \cite{Aubourg:2014yra}. The code already contains several samplers for a proper exploration of the parameter space, but in particular, we use a modified version of the nested sampler Dynesty \cite{Higson17,Speagle19} that allows us to produce posterior distributions and computes the Bayesian evidence, used to perform a model comparison through the Jeffreys scale \cite{Vazquez:2011xa}.
A Bayesian model selection was applied to get insights from cosmological functions \cite{JAVazquez12}, and in particular to the dark energy EoS in Refs. \cite{JAVazquez,Hee:2016ce,Tamayo:2019gqj}. The SimpleMC code uses a compressed version of the recent Planck CMB data (PLK), a recent reanalysis of Type Ia supernova data (SN), and high-precision BAO measurements at different redshifts up to $z=2.36$, viz., Ly-$\alpha$ DR14, BAO-Galaxy consensus, the SDSS Main Galaxy Sample (MGS) and the Six-Degree Field Galaxy Survey (6dFGS) as presented in Refs. \cite{Alam:2016hwk,Blomqvist:2019rah,Ata:2017dya,Agathe:2019vsu,Beutler2011, Anderson:2013zyy}. We do not use the final eBOSS (SDSS DR16), which contains all the data from eBOSS and its predecessor, as the covariance matrix is not available so far \cite{Alam:2020sor,duMasdesBourboux:2020pck}. We also include a collection of currently available measurements on $H(z)$ from cosmic chronometers ($H$) (see Ref. \cite{Gomez-Valent:2018hwc} and references therein). To improve the constraining power on the parameter space, we also include a big bang nucleosynthesis prior on the baryons contribution \cite{cooke}. See Ref. \cite{Padilla:2019mgi} for an extended review of the cosmological parameter inference procedure we used. In the analysis, the radiation density parameter is given by $\Omega_{\rm r0}=2.469\times10^{-5}h_0^{-2}(1+0.2271N_{\rm eff})$---where $h_0=H_0/100\, {\rm km\,s}^{-1}{\rm Mpc}^{-1}$ is the dimensionless reduced Hubble constant \cite{WMAP11a} and $N_{\rm eff}=3.046$ is the standard number of effective neutrino species with minimum allowed mass $m_{\nu}=0.06$ eV---as the present-day photon energy density is already extremely well constrained by today's CMB temperature $T_0 =2.7255 \pm 0.0006 \,{\rm K}$ \cite{Fixsen09}. Throughout our analysis, we assume flat priors over our sampling parameters: $\Omega_{{\rm m}0}=[0.05,1.0]$ for the dust density parameter today, $\Omega_{{\rm b}0} h_0^2=[0.02,0.025]$ for the physical baryon density today, and $h_0=[0.4,1.0]$ for the reduced Hubble constant. We assume $w_{\rm ci0}=[-1.5,-0.5]$ for the present-day EoS parameter of the simple-gDE \eqref{eqn:GDE} and $\Omega_{k0}=[-0.6,0.6]$ for the density parameter corresponding to the spatial curvature today.

\begin{figure*}[t!]
  \includegraphics[width=0.497\textwidth]{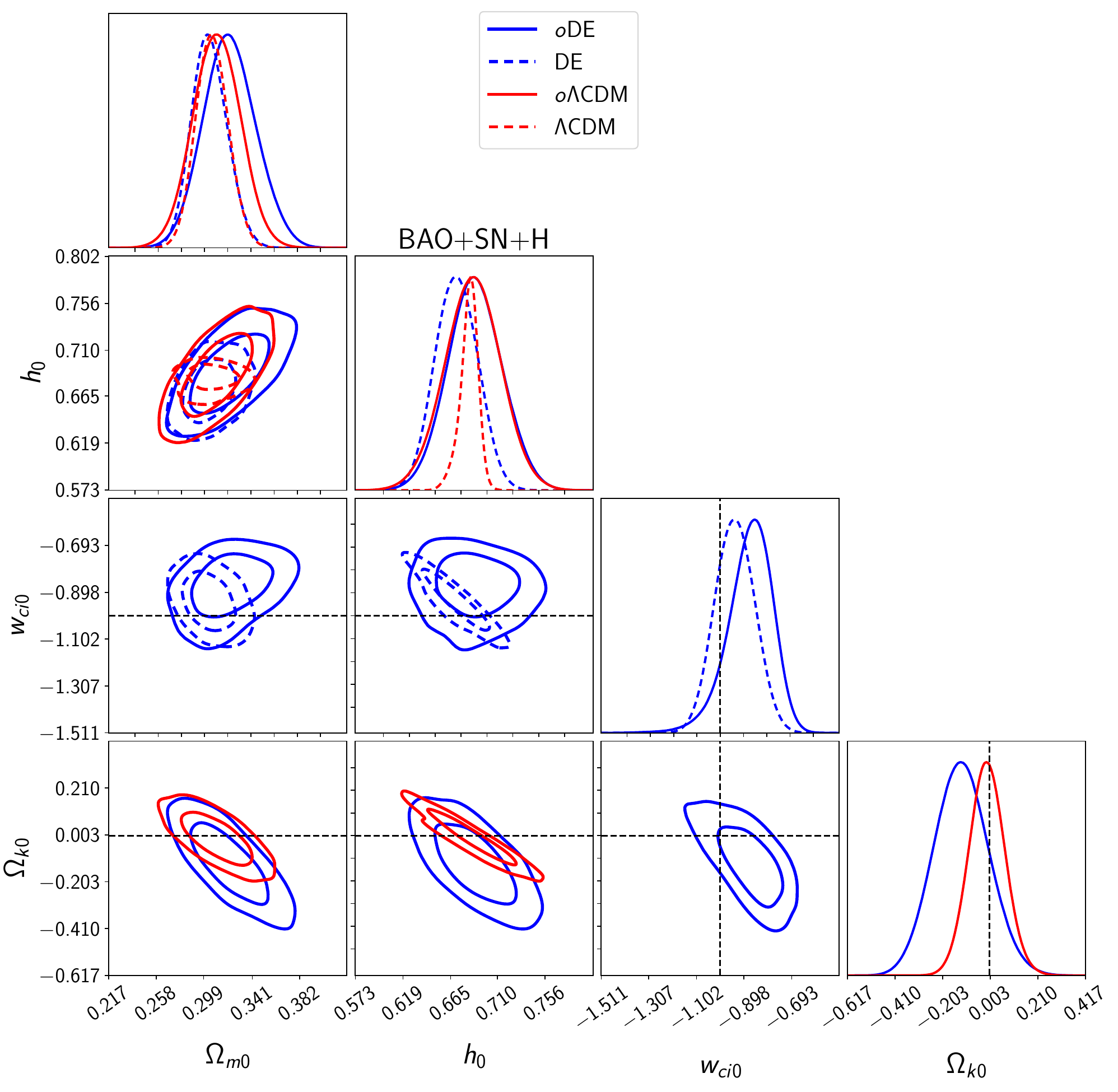}
  \includegraphics[width=0.493\textwidth]{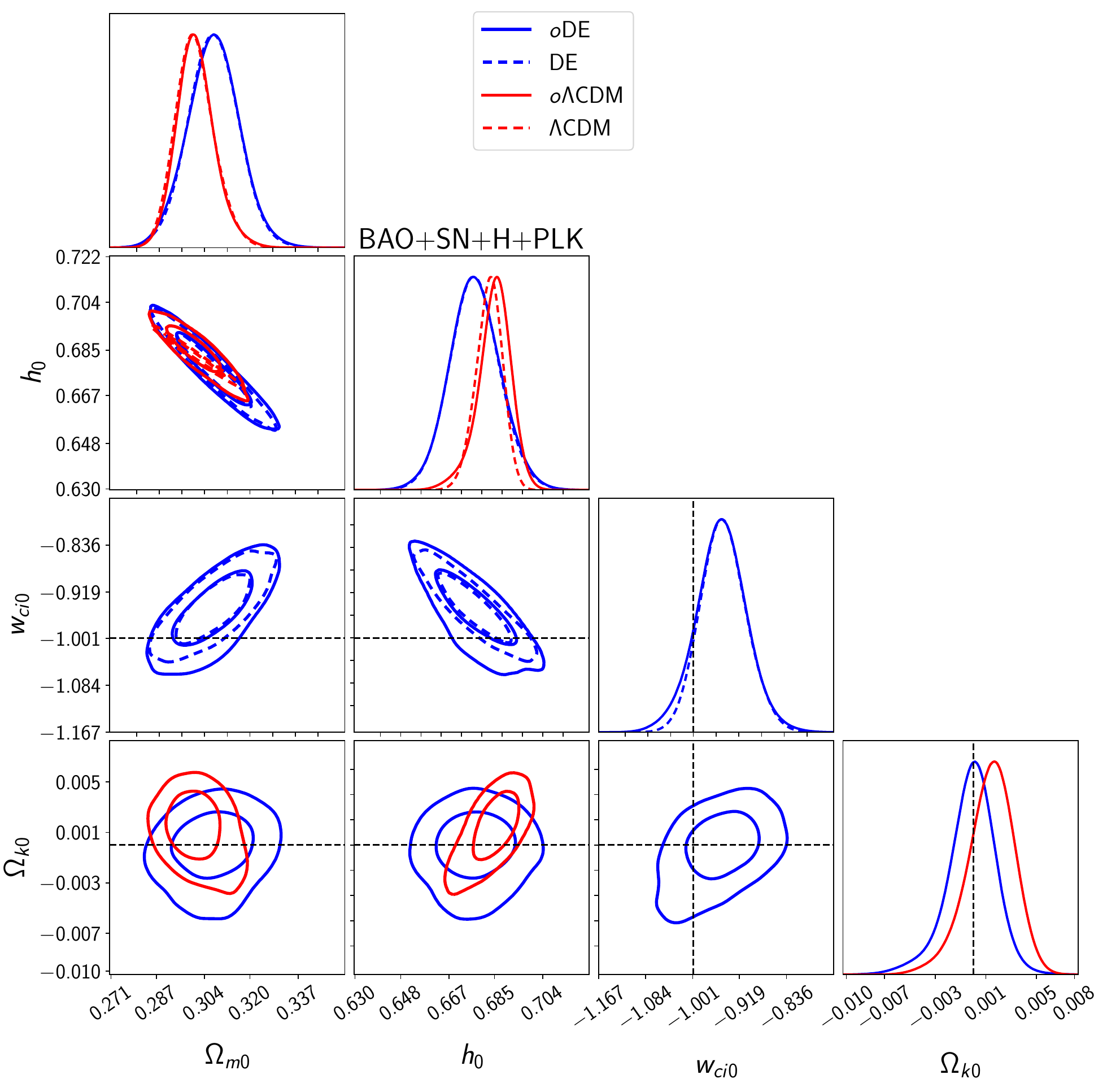}
  \caption{One- and two-dimensional ($68\%$  and  $95\%$  CLs)  marginalized posterior distributions for the free parameters of DE, $o$DE, $\Lambda$CDM, and $o\Lambda$CDM models using the combined datasets of BAO+SN+$H$ (left panel) and BAO+SN+$H$+PLK (right panel).}
  \label{fig:dy}
\end{figure*}

\begin{figure}[t!]
  \includegraphics[trim = 1mm  0mm 1mm 0mm, clip, width=0.42\textwidth]{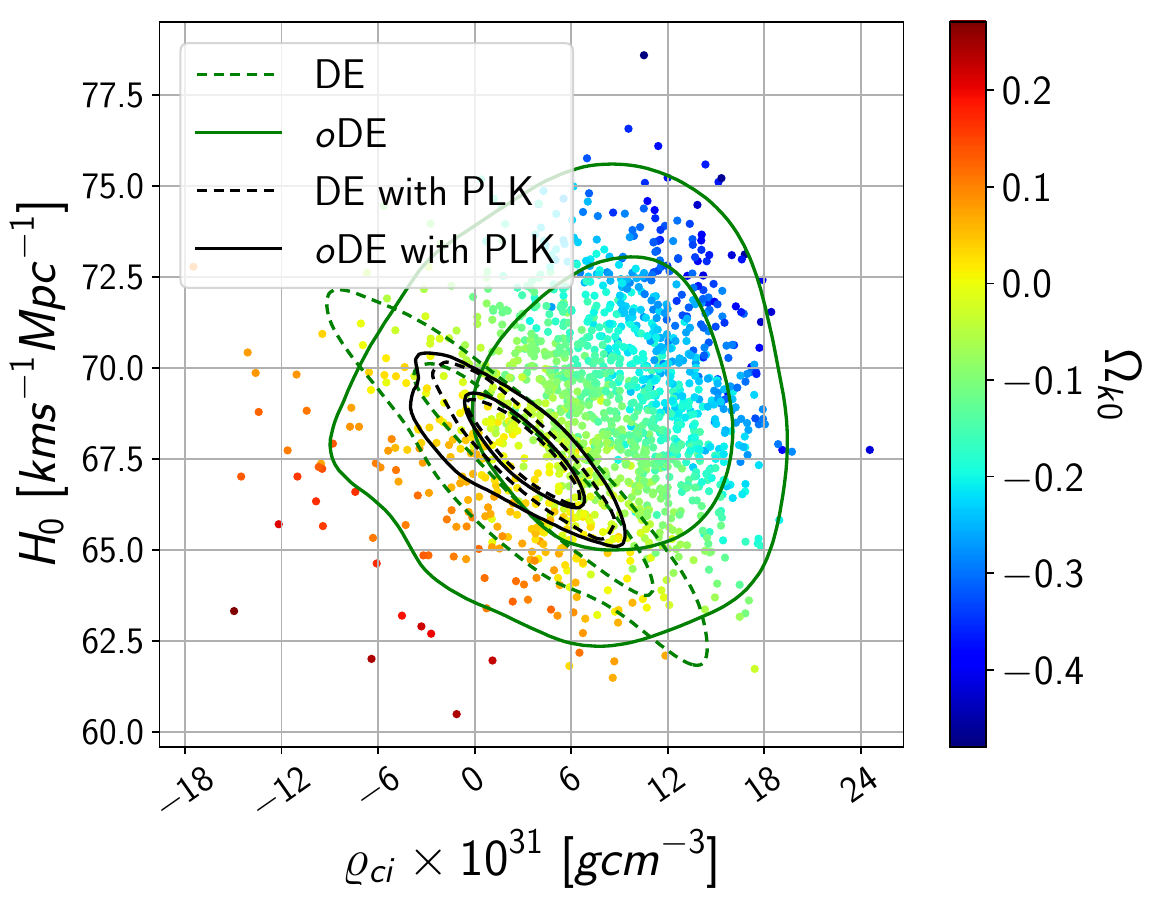}
   \caption{Two-dimensional ($68\%$  and  $95\%$  CLs)  marginalized  distributions  of $H_0$ with respect to $\varrho_{{\rm ci}}$ of the DE and of the $o$DE models for the combined BAO+SN+$H$ (without PLK) and BAO+SN+$H$+PLK (with PLK) datasets.}
   \label{fig:summary}
\end{figure}

Table~\ref{tab:priors} displays the constraints at 68\% confidence level (CL) on the free parameters---$\Omega_{{\rm m}0}$, $\Omega_{{\rm b}0}h_0^2$, $h_0$, $w_{\rm ci0}$, and $\Omega_{k0}$--- as well as the derived parameters---$\varrho_{\rm ci}$, $\Omega_{\rm ci0}$, $\Omega_{k{\rm ci}0}$, $z_{\rm ci*}$, and $z_{k{\rm ci}*}$(or $z_{k{\rm cc}*}$)--- from both the combined datasets of BAO+SN+$H$ and BAO+SN+$H$+PLK separately. In the last three rows, we list the best fit ($-2\ln{\mathcal{L}_{\rm max}}$), the $\log$-Bayesian evidence for each of the models ($\ln \mathcal{Z}$), and the $\log$-Bayesian evidence for each of the models relative to the reference model, the $\Lambda$CDM model ($\Delta\ln \mathcal{Z}=\ln \mathcal{Z}-\ln \mathcal{Z}_{{\Lambda}\rm CDM}$). According to the Jeffreys scale the Bayes factor $|\Delta\ln \mathcal{Z}|$ lying in the range [0,1) implies the strength of the evidence to be weak/inconclusive, while a positive/significant evidence is implied by the values in the range [1,3)  \cite{Jeffreys}. Figure~\ref{fig:dy}, complementary to Table~\ref{tab:priors}, displays the constraints in the form of one-dimensional marginalized posterior distributions as well as two-dimensional marginalized distributions (the inner and external contours are for $68\%$  and  $95\%$  CLs, respectively).

We start by noticing that, as is the case for both the $\Lambda$CDM and $o\Lambda$CDM models as well, the combined BAO+SN+$H$+PLK dataset puts tight constraints on the free parameters of both the DE and $o$DE models, and then the constraints on the extended models (i.e., $o\Lambda$CDM, DE, and $o$DE models) do not significantly differ from those on the $\Lambda$CDM model. Accordingly, none of the extended models suggest significant deviation in the history of the Universe up until today from the one described by the $\Lambda$CDM model, yet both the DE and $o$DE models predict futures completely different than the de Sitter future of the $\Lambda$CDM and $o\Lambda$CDM models. We find no significant improvement regarding the so-called low-redshift tensions prevailing within the $\Lambda$CDM model. Nevertheless, in case the CMB dataset is not included, i.e., the combined BAO+SN+$H$ dataset is used, in the analysis, the constraints on the extended models exhibit interesting and some significant deviations from those on the $\Lambda$CDM model that deserve a closer look. In what follows, in this section, we discuss some of the interesting results and findings along with their implications.

\begin{figure*}[t!]
  \includegraphics[trim = 2mm  4mm 35mm 2mm, clip, width=0.30\textwidth]{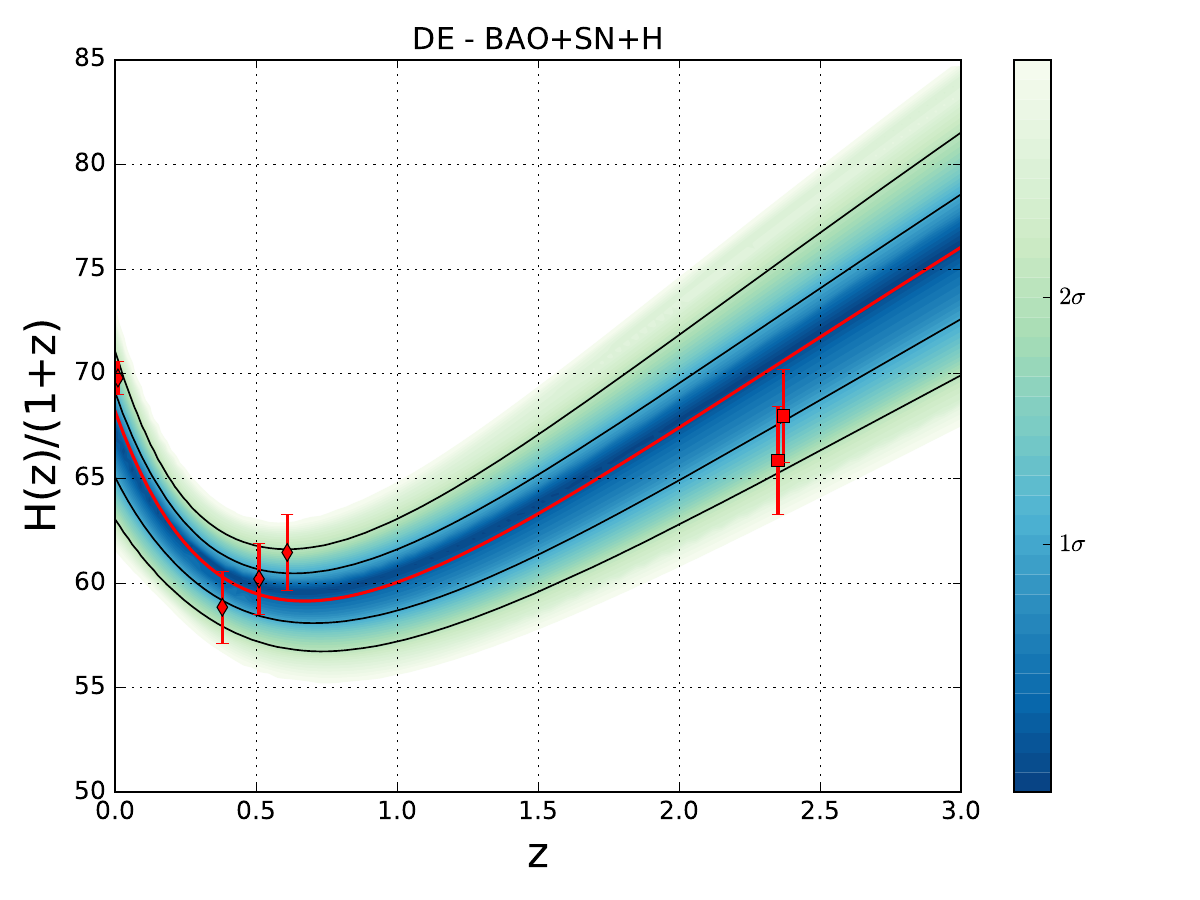}
 \includegraphics[trim = 2mm  4mm 35mm 2mm, clip, width=0.30\textwidth]{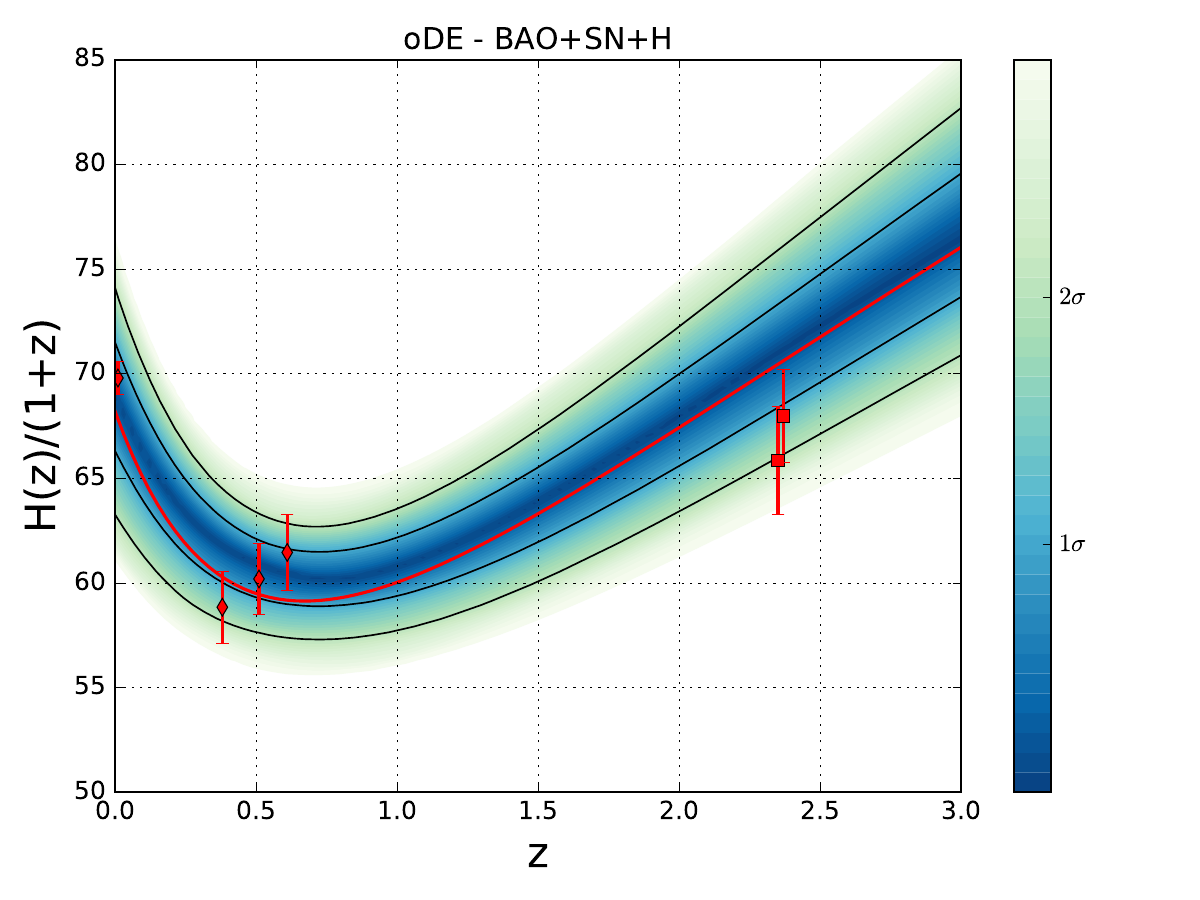}
   \includegraphics[trim = 2mm  4mm 0mm 2mm, clip, width=0.363\textwidth]{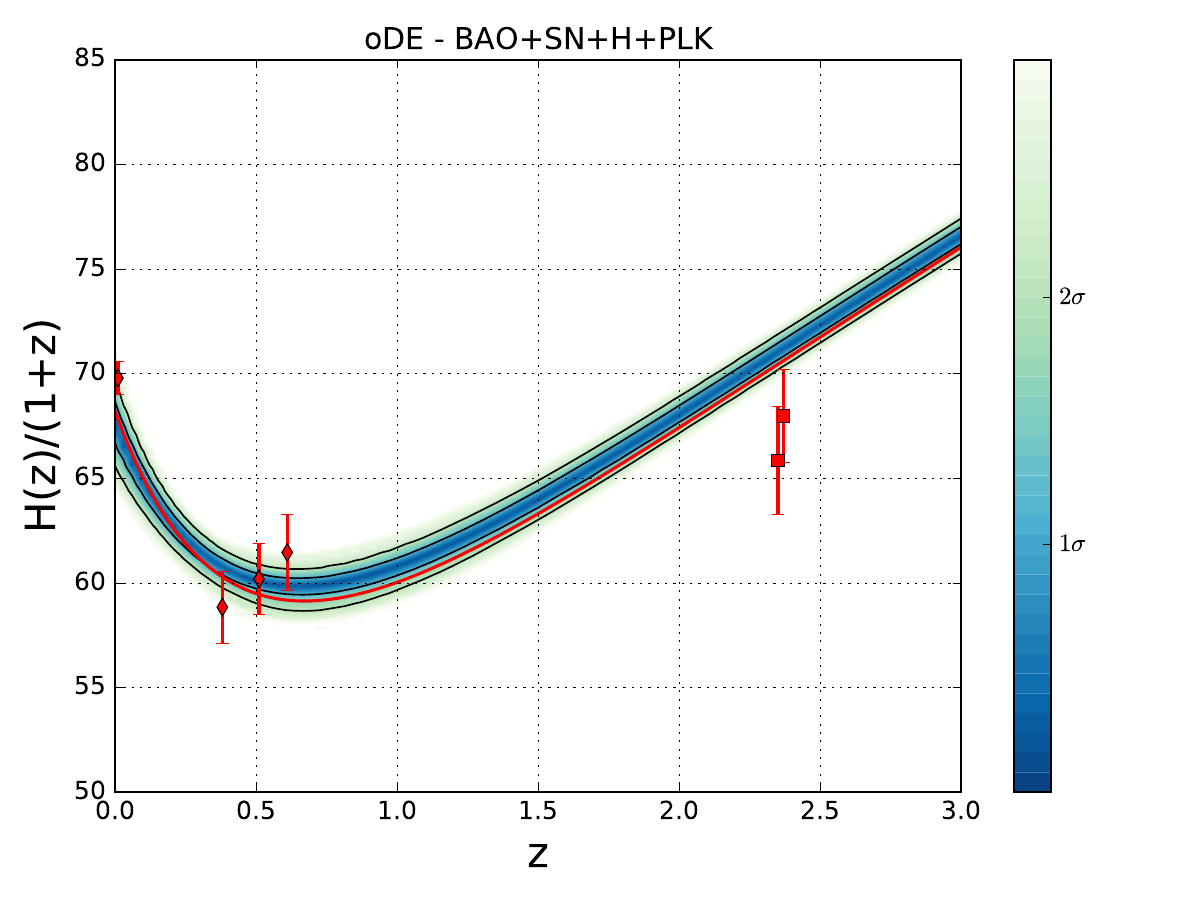}
\caption{$H(z)/(1+z)$ vs $z$ for DE (left), $o$DE (middle) models using the combined BAO+SN+$H$ datasets and for $o$DE model (right) using BAO+SN+$H$+PLK dataset, in which we consider the observational $H(z)$ values, $H_0=69.8\pm0.8\,{\rm km\,s}^{-1}{\rm Mpc}^{-1}$ from the TRGB $H_0$ \cite{Freedman:2019jwv}, BAO Galaxy consensus and Ly-$\alpha$ DR14 (red error bars) \cite{Blomqvist:2019rah,Anderson:2013zyy}.}
\label{fig:H}
\end{figure*}

\textit{In the case of BAO+SN+$H$ dataset -} 
The $o$DE model, having the lowest $-2\ln{\mathcal{L}_{\rm max}}$ value, is the one that fits best to the combined BAO+SN+$H$ dataset. The Bayesian evidence on the other hand suggests that there is a significant evidence for preferring the $\Lambda$CDM model over the extended models, as for which $|\Delta\ln \mathcal{Z}|\sim 1.5$. It is striking that, when only the models including spatial curvature are compared with each other, there is no evidence to prefer the $o\Lambda$CDM model, which yields $\Omega_{k0}=-0.011 \pm 0.077$ consistent with spatially flat universe, over the $o$DE model, which yields $\Omega_{k0}=-0.122 \pm 0.117$ suggesting spatially closed universe with high significance. The constraints on $\Omega_{\rm b0}h_0^2$ are almost exactly the same for all the models, but the constraint on the Hubble constant (or $h_0$) in the case of the DE model, $H_0=67.06\pm2.02\,{\rm km\,s}^{-1}{\rm Mpc}^{-1}$, is smaller than the one in the case of the $\Lambda$CDM model, $H_0=68.27\pm0.88\,{\rm km\,s}^{-1}{\rm Mpc}^{-1}$. Namely, the combined BAO+SN+$H$ dataset suggests that, contrary to our initial expectations discussed in Sec.~\ref{sec:DEmodel}, the simple-gDE \eqref{eqn:GDE} upgrading the null inertial mass density of the usual vacuum energy to an arbitrary constant worsens the so-called $H_0$ tension. The reason is being that the data favor $\varrho_{\rm ci}=(3.46\pm4.76)\times10^{-31}\rm g\,cm^{-3}$ (corresponding to $w_{\rm ci0}=-0.937\pm0.084$) rather than a definitely negative inertial mass destiny $\varrho_{\rm ci}<0$ (viz. phantom character today, i.e., $w_{\rm ci0}<-1$ and $\rho_{\rm ci0}>0$)---see the negative correlation between $h_0$ and $w_{\rm ci0}$ in Fig.~\ref{fig:dy}. The inclusion of spatial curvature however lifts $H_0$ to the values larger than those allowed within the $\Lambda$CDM model, as the data favor negative values of $\Omega_{k0}$---see the negative correlation between $h_0$ and $\Omega_{k0}$ in both models in Fig.~\ref{fig:dy}. In the case of the $o$DE model, the data favor spatially closed universe with high significance, viz., $\Omega_{k0}=-0.122\pm0.117$, sufficient to compensate for the decreasing effect of $w_{\rm ci0}=-0.872 \pm 0.097$ or $\varrho_{\rm ci}= (7.65\pm5.72)\times 10^{-31} \, {\rm g\, cm^{-3}}$ on the Hubble constant, and predict $H_0=68.84\pm2.60\,{\rm km\,s}^{-1}{\rm Mpc}^{-1}$. See, in Fig.~\ref{fig:dy}, the negative correlation between $\Omega_{k0}$ and $w_{\rm ci0}$, and also that the existing negative correlation between $h_0$ and $w_{\rm ci0}$ in the case of the DE model disappears with the inclusion of spatial curvature, i.e., in the case of the $o$DE model. See also Fig.~\ref{fig:summary} which demonstrates the interplay between $H_0$, $\varrho_{\rm ci}$ and $\Omega_{k0}$ in the light of the observational data. Notice that, when the $o$DE model (or the $o\Lambda$CDM model predicting $H_0=68.62\pm2.68\,{\rm km\,s}^{-1}{\rm Mpc}^{-1}$) is compared to the $\Lambda$CDM and DE models, it is the increased error along with the slightly enhanced mean value of the constraint on $H_0$ that reconciles the $o$DE model (or the $o\Lambda$CDM model) with the model independent measurements of the Hubble constant, for instance, with the distance ladder measurements, e.g., $H_0= 69.8\pm 0.8 \,{\rm km\,s}^{-1}{\rm Mpc}^{-1}$ from a recent calibration of the tip of the red giant branch (TRGB) applied to Type Ia supernovae \cite{Freedman:2019jwv}. Thus, neither the $o\Lambda$CDM model nor the $o$DE model can robustly address the so-called $H_0$ problem. Similarly, the $o$DE model, as well as the DE model, better agrees with the Ly-$\alpha$ data from $z\approx 2.34$ due to the widening in the 1D marginalized posterior distribution of $H_0$, not due to a non-monotonic evolution of $H(z)$ at about $z\sim1$ as a result of a $\rho_{k\rm ci}$ assuming negative density values in the past. See Fig.~\ref{fig:H}, wherein we present the probability distribution of the $H(z)/(1+z)$ vs $z$ in the case of the combined BAO+SN+$H$ dataset for the DE (left) and $o$DE (middle) models and include the  TRGB $H_0$ and BAO data points for comparison (solid lines represent 1$\sigma$ and 2$\sigma$ accordingly and darker implies more probable as shown in the color bar).

\begin{figure}[t!]
  \includegraphics[trim = 1mm  0mm 1mm 0mm, clip, width=0.4\textwidth]{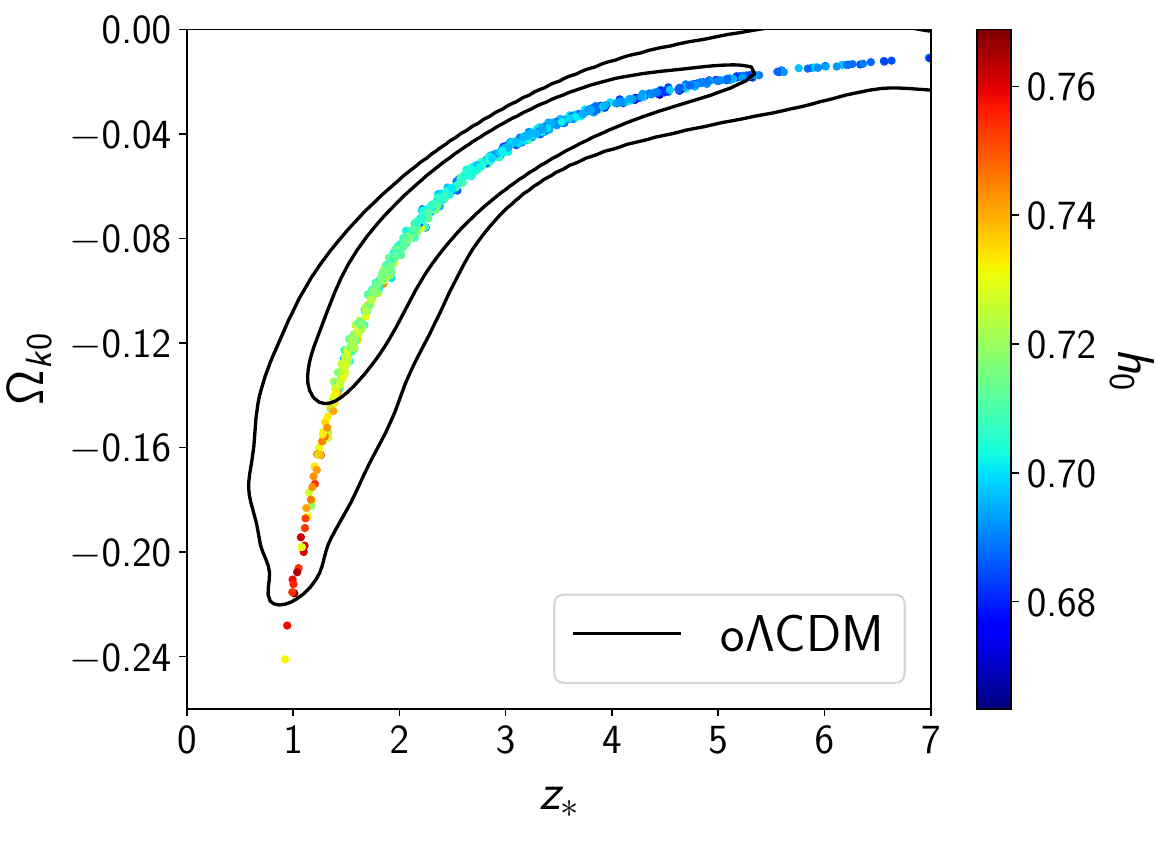}
\includegraphics[trim = 1mm  0mm 1mm 0mm,  clip, width=0.4\textwidth]{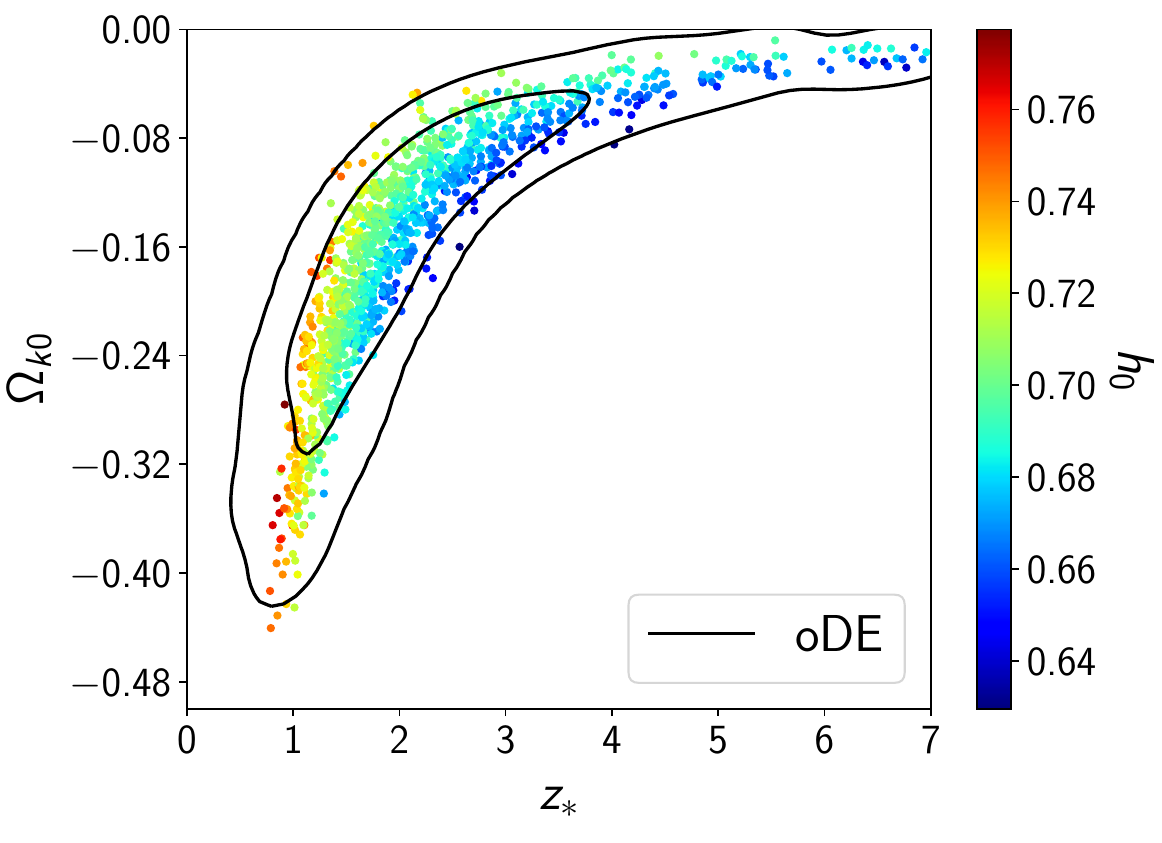}
  \caption{Two-dimensional ($68\%$  and  $95\%$  CLs)  marginalized  distributions  of  the  $h_0$, $\Omega_{k0}$ with respect to $z_{k{\rm ci}*}$ of $o$DE model and to $z_{k{\rm cc}*}$ of $o\Lambda$CDM for the combined BAO+SN+$H$ dataset.}
 \label{fig:2d2d}
\end{figure}

We observe that the reason $H_0$ takes larger values is that, as we go from today ($z=0$) to the past, the energy density of the effective source $\rho_{k{\rm ci}}$ \eqref{eq:rhotilde} can cross below zero in the recent past, at a redshift $z_{k \rm ci*}>0.92$ for the $o$DE model and $z_{k \rm ci*}>1.26$ for the $o\Lambda$CDM model. In both models, this happens because of the closed space ($\Omega_{k0}<0$), whereas the simple-gDE opposes it---notice that the energy density of the simple-gDE never crosses below zero in the past, but in the far future ($z_{\rm ci*}<-0.78$). In the DE model, in which $H_0$ takes the smallest values among all the models, the simple-gDE cannot cross below zero in the recent past, but either in the far future ($z_{\rm ci*}<-0.96$) or remote past ($z_{\rm ci*}\gtrsim10^7$). The interplay between $\Omega_{ k0}$, $h_0$ (or $H_0$), and $z_{k\rm ci*}$, in the light of observational data, is well demonstrated in Fig.~\ref{fig:2d2d}: The larger negative values of $\Omega_{k0}$, the lower the value of $z_*$ (viz., $z_{k\rm ci*}$ for the $o$DE model and $z_{k\rm cc*}$ for the $o\Lambda$CDM model) accompanying by larger (more red) values of $h_0$ (or $H_0$). The Hubble constant achieves its largest values within 68\% CL,  $H_0\sim 72\,{\rm km\,s}^{-1}{\rm Mpc}^{-1}$, if $z_*\sim1$ (the smallest $z_*$ value allowed within 68\% CL). On the other hand, for large $z_*$ values, the $H_0$ values predicted within the $o\Lambda$CDM and $o$DE models approach those predicted within the $\Lambda$CDM and DE models, respectively.

%Thus minimal deviation from the cosmological constant does not help employing the mechanism suggested in the beginning but instead opposes it. This may be interpreted as that a source that could help this mechanism to work would not be a simple deviation from the cosmological constant (See \cite{Akarsu:2019hmw} for a relevant extended discussion). However, something interesting is going on here: The compensation of the EoS $w_{\rm ci0}$ very much above -1 (implying increased energy density w.r.t the cosmological constant at large redshifts) results in the larger negative values of $\Omega_{k0}$. The data seek for negative energy densities, provided by the spatial curvature, as the dark energy deviates away from $\Lambda$ towards EoS larger than minus unity to compensate it the data need even larger negative $\Omega_{k0}$

\textit{In the case of BAO+SN+$H$+PLK dataset -}
The DE and $o$DE models fit to the combined BAO+SN+$H$+PLK equally well (having the same $-2\ln{\mathcal{L}_{\rm max}}$ value), and both fit better than the $\Lambda$CDM and $o\Lambda$CDM models. The Bayesian evidence, on the other hand, presents significant evidence against the models including spatial curvature, as it is $|\Delta\ln \mathcal{Z}|\sim 2$ for both the $o\Lambda$CDM and $o$DE models, and suggests there is no evidence to prefer the $o\Lambda$CDM model over the DE model, as it is $\Delta\ln \mathcal{Z}=-0.17\pm0.36$ for the DE model. The inclusion of the Planck CMB data in the analysis does not considerably change the constraints on the parameters of the $o\Lambda$CDM model, though $\Omega_{k0}$ now favors slightly positive values, but still remains consistent with spatially flat universe, viz., $\Omega_{k0}=0.0012\pm0.0018$. Nevertheless, the constraints on all the common parameters of the $\Lambda$CDM and $o\Lambda$CDM models now become almost the same. On the other hand, the inclusion of the Planck CMB data in the analysis considerably changes the constraints on the parameters of the $o$DE model---in particular, on the parameters $\Omega_{k0}$ and $w_{\rm ci0}$--- and brings by predicting $\Omega_{k0}=-0.0001\pm0.0019$, the $o$DE model almost indistinguishably close to the DE model. It is remarkable that the $o$DE model now prefers spatial flatness with a precision higher than the $o\Lambda$CDM model does; in spite of that, when the Planck CMB data were not included in our analyses, the $o\Lambda$CDM model was consistent with spatial flatness, but the $o$DE model was not. This resurrection of the spatial flatness is accompanied by the fact that the simple-gDE now resembles the usual vacuum energy, namely, it still yields positive inertial mass density, but now closer to zero, $\varrho_{\rm ci}= (2.85\pm2.58)\times 10^{-31} \, {\rm g\, cm^{-3}}$. This constraint now is almost the same as the one in the case of the DE model, $\varrho_{\rm ci}= (3.06\pm2.28)\times 10^{-31} \, {\rm g\, cm^{-3}}$, whereas it was allowed to take negative values when the Planck CMB data were not included in our analysis; see Fig.~\ref{fig:summary}. These slightly positive inertial mass density values correspond to the present-day EoS parameters of the simple-gDE $w_{\rm ci0}=-0.948\pm0.041$ for the DE model and $w_{\rm ci0}=-0.951\pm0.045$ for the $o$DE model and then result in slightly smaller $H_0$ values, $H_0=67.72\pm0.97\,{\rm km\,s}^{-1}{\rm Mpc}^{-1}$ for the DE model and $H_0=67.73\pm0.99\,{\rm km\,s}^{-1}{\rm Mpc}^{-1}$ for the $o$DE model, compared to the standard $\Lambda$CDM value $H_0=68.29\pm0.52\,{\rm km\,s}^{-1}{\rm Mpc}^{-1}$---see the negative correlation between $h_0$ and $w_{\rm ci0}$ in Fig.~\ref{fig:dy} as well as the negative correlation between $H_0$ and $\varrho_{\rm ci}$ in Fig.~\ref{fig:summary}. In contrast to the results without the Planck CMB data, the deficit in $H_0$ in the case of the $o$DE model cannot be compensated by the spatial closedness of the universe, as with the inclusion of the Planck data the present universe is spatially flat with an accuracy of 0.2\%. The data predict $z_{k\rm cc*}\sim10$ ($o\Lambda$CDM) and $z_{k\rm ci*}\sim7$ ($o$DE) for the lower limit of the redshift at which the energy density of the effective source crosses below zero in the $o\Lambda$CDM and $o$DE models. And, even these lower limits remain too large for the energy density crossing below zero to be efficient in enhancing the constraints on $H_0$ in the $o\Lambda$CDM and $o$DE models.

Finally, one may see Fig.~\ref{fig:summary} for a very good summary of the observational constraints on the two minimal extensions of the standard $\Lambda$CDM model---(i) the spatial curvature ($\Omega_{k0}$) and (ii) simple-gDE promoting the null inertial mass density of the usual vacuum energy to an arbitrary constant $\varrho_{\rm ci}$---and their influence on the Hubble constant $H_0$. The combined BAO+SN+$H$+PLK dataset presents evidence at equal strength for the usual vacuum energy (null inertial mass density) and the simple-gDE with a constant inertial mass density very close to zero, viz., at the order of $\mathcal{O}(10^{-12})\,\rm eV^4$. This latter possibility predicts almost exactly the same history of the universe up until today as the standard $\Lambda$CDM model---so that it does not result in any improvement regarding the Ly-$\alpha$ BAO measurements--- except that it slightly aggravates the so-called $H_0$ tension prevailing within the standard $\Lambda$CDM model; see the right panel in Fig.~\ref{fig:H}. Yet, the constraints on the model considering simple-gDE instead of the $\Lambda$ suggest totally different futures for the universe in comparison to the standard $\Lambda$CDM model. As we discussed in Sec.~\ref{sec:DEmodel}, the future will be drastically different depending on the sign of $\varrho_{\rm ci}$: a bouncing ($H=0$) universe in the finite future if $\varrho_{\rm ci}>0$ and a forever expanding universe for $\varrho_{\rm ci}\leq0$ with the infinite future limit of the de Sitter Universe for $\varrho_{\rm ci}=0$ and of the LSBR for $\varrho_{\rm ci}<0$. In the next section, in light of the observational analyses presented here, we will discuss in detail the complete history of the Universe predicted by these four models under consideration in this paper.

\section{Dynamical Analysis}
\label{sec:dynamicalanalysis}
 To analyze the asymptotic behaviour of the models under consideration, we resort to the well-known methods of dynamical systems. In the case of homogeneous cosmologies such as the one analyzed here, one usually starts by defining a set of dimensionless variables, e.g., the density parameters introduced above, and then proceeds to study their evolution in the parameter space by transforming Einstein and conservation equations in terms of the new variables. The fixed (or critical) points of such a system represent classes of solutions that can be interpreted as cosmological phases with a specific matter component dominating. In the present context, such a method allows us to determine under which conditions the scenarios envisioned by the DE and $o$DE models can be considered as future phases of our Universe (i.e., future attractors of the dynamical system). This will be done by implementing the constraints on the free parameters from both the combined datasets of BAO+SN+$H$ and BAO+SN+$H$+PLK presented in Table~\ref{tab:priors}.  

Cosmologies with spatial curvature or simple-gDE different from the $\Lambda$CDM model can present recollapsing as well as bouncing scenarios: in such cases, the usual expansion-normalized variables are ill defined due to the vanishing of the expansion $H$ at the turning points of the scale factor. With the aim of capturing also these cases in a global stability analysis, we construct properly the dimensionless autonomous system and define the following dimensionless variables,
\begin{align}
 X_{\rm m}=\frac{\rho_{\rm m}}{3D^2}\,,\,\, X_{\rm r}=\frac{\rho_{\rm r}}{3D^2}\,,\,\,X_{\rm ci} = \frac{\rho_{\rm ci0}}{3D^2}\,,\,\,X_{H} = \frac{H}{D},\label{var1}
\end{align}
where the normalization $D$ and its evolution are governed by 
\begin{equation}
 D^2 = H^2 + \frac{|k|}{a^2}\quad ,\quad  \frac{\dot{D}}{D^2} = X_{H} \left( \frac{\dot{H}}{D^2} + X_{H}^2 -1 \right).
\end{equation}
By inspecting the dimensionless variables above, we see that $X_{\rm m}$ and $X_{\rm r}$ are positive definite. The sign of $X_{\rm ci}$ depends on the the sign of $\rho_{\rm ci0}$, which has already been supposed to be positive. Finally, $X_{H}$ is defined in the interval $[-1,1]$ and its sign depends on the sign of the Hubble parameter $H$; it is positive for expanding models and negative for collapsing ones. The boundary values $+1$ and $-1$ of this variable correspond to spatially flat expanding and collapsing models, respectively. The normalization $D$ has been introduced in Ref. \cite{Goliath:1998na} as a means to compactify the parameter space of spatially homogeneous cosmologies and include bouncing/recollapsing scenarios in the analysis, as is well defined throughout the whole cosmological evolution including possible turning points of the scale factor (see also Ref. \cite{Bahamonde:2018} for a comprehensive review of this and other methods to treat noncompact dynamical systems). To decouple the dynamics of $D$ from the dynamics of the other variables, we define the new time parameter ${\rm d}\tau = D\, {\rm d}t$ and take derivatives of the definitions given in \eqref{var1} with respect to $\tau$:
\begin{align}
 X_{\rm m}' &= -2\, X_{\rm m}\, X_{H}\, \left( \frac{\dot{H}}{D^2} + X_{H}^2 +\frac{1}{2} \right), \label{eq:Xmp}\\
 X_{\rm r}' &= -2\, X_{\rm r}\, X_{H}\, \left( \frac{\dot{H}}{D^2} + X_{H}^2 + 1 \right), \label{eq:Xrp}\\
 X_{\rm ci}' &= -2\, X_{\rm ci}\, X_{H}\, \left( \frac{\dot{H}}{D^2} + X_{H}^2 -1 \right), \label{eq:Xde0p}\\
 X_{H}' &= \left( 1 - X_{H}^2 \right)\, \left( \frac{\dot{H}}{D^2} + X_{H}^2 \right). \label{eq:Xhp}
\end{align}
The fixed points of such a system describe specific asymptotic cosmological solutions, whose stability in general depends on the free parameters involved. The cosmological interpretation of the critical points is expressed in terms of the effective EoS parameter ($w_{\rm eff} \equiv p_{\rm eff}/\rho_{\rm eff}$)
\begin{align}\label{eos_gen}
 w_{\rm eff} = \frac{\frac{1}{3}X_{\rm r}+X_{\rm ci}\, \left[w_{\rm ci0}+3(1+w_{\rm ci0})\, \ln a\right]}{X_{\rm m}+X_{\rm r}+X_{\rm ci}\, \left[1-3(1+w_{\rm ci0})\, \ln a \right]},
\end{align}
and deceleration parameter
\begin{align}\label{decel_gen}
 q &= -1-\frac{1}{X_{H}^{2}}\frac{\dot{H}}{D^2}.
\end{align}
The $\dot{H}$ and $\ln a$ terms in all the previous equations will be provided by Raychaudhuri and Friedmann equations, and they will have different forms in terms of the dimensionless variables depending on the sign of the spatial curvature (viz., $k$); hence, we are going to present the analysis of the critical points of the system for the cases $k\leq 0$ ($\Omega_{k0}\geq 0$) and $k>0$  ($\Omega_{k0}<0$) separately. 
\subsection{Nonpositive spatial curvature $k\leq0$ ($\Omega_{k0}\geq 0$)}
\label{sec:negcurv}
For nonpositive spatial curvature (flat/open space), Friedmann and Raychaudhuri equations in terms of the dimensionless variables take the following forms, respectively:
\begin{align}
 &2\, X_{H}^2 - 1 =X_{\rm m} + X_{\rm r} +\left[ 1-3\, (1+w_{\rm ci0})\, \ln a \right]X_{\rm ci}, \label{eq:friedneg}\\
  &\frac{\dot{H}}{D^2}=\frac{1}{2}-2X_{H}^2-\frac{1}{2}X_{\rm r}-\frac{3}{2}\left[ w_{\rm ci0}+3\, (1+w_{\rm ci0})\, \ln a \right]X_{\rm ci}.
 \end{align}
By substituting $\ln a$ from the Friedmann equation into the Raychaudhuri equation, we obtain
%\begin{equation}\label{eq:rayneg}
% \frac{\dot{H}}{D^2} = %\frac{1}{2}\left[ 2\, X_{H}^2-2-3\, %X_{\rm m}-4\, X_{\rm r}-3\, (1+w_{\rm %ci0})\, X_{\rm ci} \right].
%\end{equation}
\begin{equation}\label{eq:rayneg}
 \frac{\dot{H}}{D^2} =-1+X_{H}^2-\frac{3}{2}\, X_{\rm m}-2\, X_{\rm r}-\frac{3}{2}\, (1+w_{\rm ci0})\, X_{\rm ci}.
\end{equation}
This expression closes the autonomous system in the case of negative spatial curvature.  The effective EoS and deceleration parameters are given by
\begin{align}
 w_{\rm eff} &= -1+\frac{3\, X_{\rm m}+4\, X_{\rm r}+3\, (1+w_{\rm ci0})\, X_{\rm ci}}{3\, (1-2\, X_{H}^2)},\\
 q &= \frac{2+3\, X_{\rm m}+4\, X_{\rm r}+3\, (1+w_{\rm ci0})\, X_{\rm ci}}{2\, X_{H}^2}-2\, . 
\end{align}

\subsection{Positive spatial curvature $k>0$ ($\Omega_{k0}<0$)}
\label{sec:poscurv}
For positive spatial curvature (closed space), Friedmann and Rauchaudhuri equations are
%\begin{align}
% 1&=X_{\rm m} + X_{\rm r} + X_{\rm ci}\, [ 1-3\, (1+w_{\rm ci0})\, \ln a ], \label{eq:friedpos} \\
% \frac{\dot{H}}{D^2}&=-\frac{1}{2}\{1+2\, X_{H}^2+X_{\rm r}+3\, X_{\rm ci}\, \left( w_{\rm ci0}+3\, (1+w_{\rm ci0})\, \ln a \right) \}\, .
%\end{align}
\begin{align}
 1&=X_{\rm m} + X_{\rm r} + X_{\rm ci}\, \left[ 1-3\, (1+w_{\rm ci0})\, \ln a \right], \label{eq:friedpos} \\
 \frac{\dot{H}}{D^2}&=-\frac{1}{2}-\, X_{H}^2-\frac{1}{2}X_{\rm r}-\frac{3}{2}\left[ w_{\rm ci0}+3\, (1+w_{\rm ci0})\, \ln a \right]X_{\rm ci}\, .
\end{align}
Following the same procedure as before, we substitute $\ln a$ from the Friedmann equation into the Raychaudhuri equation to obtain
%\begin{equation}
%\label{eq:raypos}
% \frac{\dot{H}}{D^2} = %\frac{1}{2}\left[ 2-2\, X_{H}^2-3\, %X_{\rm m}-4\, X_{\rm r}-3\, (1+w_{\rm %ci0})\, X_{\rm ci} \right]\, ,
%\end{equation}
\begin{equation}
\label{eq:raypos}
 \frac{\dot{H}}{D^2} = 1-X_{H}^2-\frac{3}{2}\, X_{\rm m}-2\, X_{\rm r}-\frac{3}{2}\, (1+w_{\rm ci0})\, X_{\rm ci}\, ,
\end{equation}
giving the closure of the autonomous system for positive spatial curvature.  The cosmological parameters in this case can be calculated through the following expressions:
\begin{align}
 w_{\rm eff} &= -1+X_{\rm m}+\frac{4}{3}\, X_{\rm r}+(1+w_{\rm ci0})\, X_{\rm ci},\\
 q &= \frac{3\, X_{\rm m}+4\, X_{\rm r}+3\, (1+w_{\rm ci0})\, X_{\rm ci}-2}{2\, X_{H}^2}\, .
\end{align}

\subsection{Critical points}

The critical points of the system are found by solving $\mathbf{X}'=0$ given by \eqref{eq:Xmp}--\eqref{eq:Xhp}, together with \eqref{eq:rayneg} for $k\leq0$ or \eqref{eq:raypos} for $k>0$. The coordinates of the critical elements in the parameter space are listed in Table~\ref{table:1a} together with the eigenvalues of their stability matrix and their stability character, for the case $X_{\rm ci}>0$.  We note that the dynamical system is invariant under the simultaneous change of sign $\{\, (1+w_{\rm ci0})\rightarrow-(1+w_{\rm ci0})\, ,\, X_{\rm ci}\rightarrow-X_{\rm ci}\, \}$, because these quantities appear always in the combination $(1+w_{\rm ci0}) X_{\rm ci}$.  Consequently, if one is interested in analyzing the case $w_{\rm ci0}<-1$ with positive dark energy density, the results will be the same as in the case $w_{\rm ci0}>-1$ with negative dark energy density.

\begin{table*}[ht!]
\caption{Coordinates, eigenvalues and stability of the critical points (CPs) of the system, assuming $X_{\rm ci}>0$.}
\label{table:1a}
\centering 
\resizebox{15cm}{!}{
 \begin{tabular}{|c|c|c|c|c|c|c|c|}
  \hline
  $k$ & CP & $X_{\rm m}$ & $X_{\rm r}$ & $X_{\rm ci}$ & $X_{H}$ & Eigenvalues & Stability\\[0.1cm] 
 \hline\hline
 \multirow{5}{*}{\vspace{-3.7cm}$k\leq0$} & $A_+$ & 0 & 0 & 0 & $1$ & $\{ -4 , -3 , -2 , 0 \}$ &  $\Bigg{\{}$ \pbox[c][1.5cm]{2.5cm}{\text{sink} $w_{\rm ci0}<-1$ \\ \text{saddle} $w_{\rm ci0}>-1$} \\[0.2cm] 
  & $A_-$ & 0 & 0 & 0 & $-1$ & $\{ 4 , 3 , 2 , 0 \}$ & $\Bigg{\{}$\pbox[c][1.5cm]{2.5cm}{\text{source} $w_{\rm ci0}<-1$ \\ \text{saddle} $w_{\rm ci0}>-1$}\\[0.2cm]
  & $B_+$ & 1 & 0 & 0 & $1$ & $\{ 3, 3,-1,1 \}$ & saddle\\[0.2cm]
  & $B_-$ & 1 & 0 & 0 & $-1$ & $\{ -3,-3,-1,1 \}$ & saddle\\[0.2cm]
  & $C_+$ & 0 & 1 & 0 & $1$ & $\{ 4,4,2,1 \}$ & source\\[0.2cm]
  & $C_-$ & 0 & 1 & 0 & $-1$ & $\{ -4,-4,-2,-1 \}$ & sink\\[0.2cm]
  & $D_+$ & 0 & 0 & 0 & $1/\sqrt{2}$ & $\{ -\sqrt{2},\sqrt{2},\sqrt{2},1/\sqrt{2} \}$ & saddle\\[0.2cm]
  & $D_-$ & 0 & 0 & 0 & $-1/\sqrt{2}$ & $\{ -\sqrt{2},-\sqrt{2},\sqrt{2},-1/\sqrt{2} \}$ & saddle\\[0.2cm]
  & $E$ & $-\frac{1}{3}\left( 2+4\, X_{\rm r}+3\, (1+w_{\rm ci0})\, X_{\rm ci} \right)$ & $\forall$ & $\forall$ & 0 & $\left\{ 0 ,0, -\sqrt{2 X_{\rm r}-\frac{9}{2}(1+w_{\rm ci0}) X_{\rm ci} -1 }, \sqrt{2 X_{\rm r}-\frac{9}{2}(1+w_{\rm ci0}) X_{\rm ci} -1 } \right\}$ & saddle\\[0.3cm]
 \hline
 $k>0$ & $F$ & $\frac{1}{3}\left( 2-4\, X_{\rm r}-3\, (1+w_{\rm ci0})\, X_{\rm ci} \right)$ & $\forall$ & $\forall$ & 0 & $\left\{ 0 ,0, -\sqrt{2 X_{\rm r}-\frac{9}{2}(1+w_{\rm ci0}) X_{\rm ci} +1 }, \sqrt{2 X_{\rm r}-\frac{9}{2}(1+w_{\rm ci0}) X_{\rm ci} +1 } \right\}$ & center\\ [0.2cm]
 \hline
\end{tabular}}
\end{table*}

There can be up to ten critical points in the phase space. In what follows, we go through each critical point discussing its mathematical and physical features:
\begin{enumerate}[label=(\roman*)]
    \item Points $A_\pm$: They describe exponentially expanding ($A_+$) or collapsing ($A_-$) spatially flat models, with $w_{\rm eff}=-1$ and $q=-1$. A vanishing eigenvalue signals that the system is {\it indifferently unstable} along the $X_{H}$ eigendirection. A numerical study tells us that the stability along the eigendirection depends on $w_{\rm ci0}$: assuming $X_{\rm ci}>0$, the expanding model is a future attractor if $w_{\rm ci0}<-1$, while it is an unstable saddle if $w_{\rm ci0}>-1$; the collapsing model instead is a past source if $w_{\rm ci0}<-1$ and an unstable saddle if $w_{\rm ci0}>-1$.
    \item Points $B_\pm$: They are spatially flat dust-dominated solutions, with $w_{\rm eff}=0$  and $q=1/2$.  Such points are always unstable saddles.
    \item Points $C_\pm$: They are spatially flat radiation-dominated models with $w_{\rm eff}=1/3$ and $q=1$.  The expanding solution $C_+$ is a past source, while the contracting $C_-$ is a future attractor.
    \item Points $D_\pm$: These points represent the Milne Universe with scale factor $a \sim t$. They exist only in the case of negative spatial curvature (open space), and they are unstable saddles. 
    \item Set E: The set of points $E$ is a static model ($X_{H}=0$) that can be sourced by different combinations of parameters satisfying the relation
    %\begin{equation}\label{eq:pointe}
    %X_{\rm m} = -\frac{1}{3}[ 2+4\, X_{\rm r}+3\, (1+w_{\rm ci0})\, X_{\rm ci} ].
    %\end{equation}
      \begin{equation}\label{eq:pointe}
    X_{\rm m} =-\frac{2}{3}-\frac{4}{3}\, X_{\rm r}-\, (1+w_{\rm ci0})\, X_{\rm ci}.
    \end{equation}
    Such a model has effective EoS parameter $w_{\rm eff}=-1/3$. Plugging the coordinates of the point back into Friedmann \eqref{eq:friedneg}, one can obtain the scale factor of the static universe:
    \begin{equation}\label{eq:scaleE}
     a = \exp{\left( \frac{1}{3(1+w_{\rm ci0})}+\frac{1-X_{\rm m}-X_{\rm r}}{2+3\, X_{\rm m}+4\, X_{\rm r}} \right)}.
    \end{equation}
    From \eqref{eq:pointe}, assuming the physically reasonable requirements that $X_{\rm m}\geq0$ and $X_{\rm r}\geq0$, together with the assumption $X_{\rm ci}>0$, then we must have $w_{\rm ci0}<-1$  (or $X_{\rm ci}<0$ and $w_{\rm ci0}>-1$).  For instance, in the case of purely simple-gDE-dominated universe ($X_{\rm m}=X_{\rm r}=0$), in order to keep the model static, one needs $X_{\rm ci}=-\frac{2}{3\, (1+w_{\rm ci0})}$. Owing to the presence of $w_{\rm ci0}$, the solution corresponding to this critical point does not exist in the case of the usual vacuum energy ($w_{\rm ci0}\rightarrow-1$), where a static universe is possible only for positive spatial curvature (closed  space).  According to the ranges of values specified after \eqref{eq:scaleE}, the set is unstable.

\begin{figure}[t!]
 \includegraphics[width=0.35\textwidth]{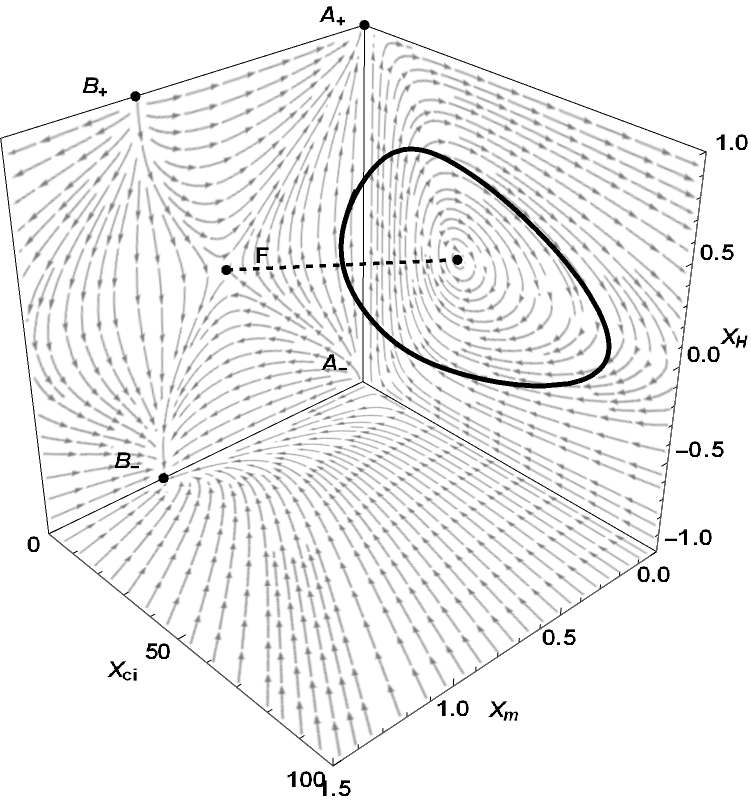}
  \caption{Example of trajectory circling around the critical line $F$ in the invariant subset $X_{\rm r}=0$.  Initial conditions are not related to observations and are chosen in such a way to show the existence of cyclic dynamics in the $o$DE model.}
  \label{fig:3d_perio}
\end{figure}

\begin{figure*}[ht!]
  \includegraphics[width=0.3\textwidth]{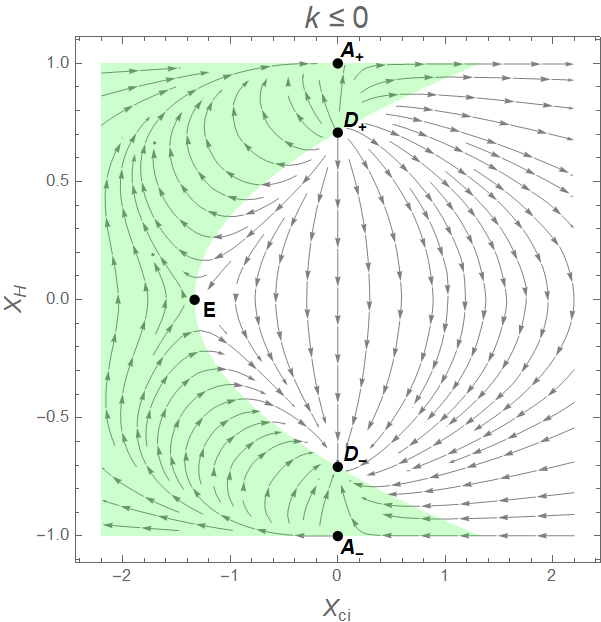}
  \includegraphics[width=0.3\textwidth]{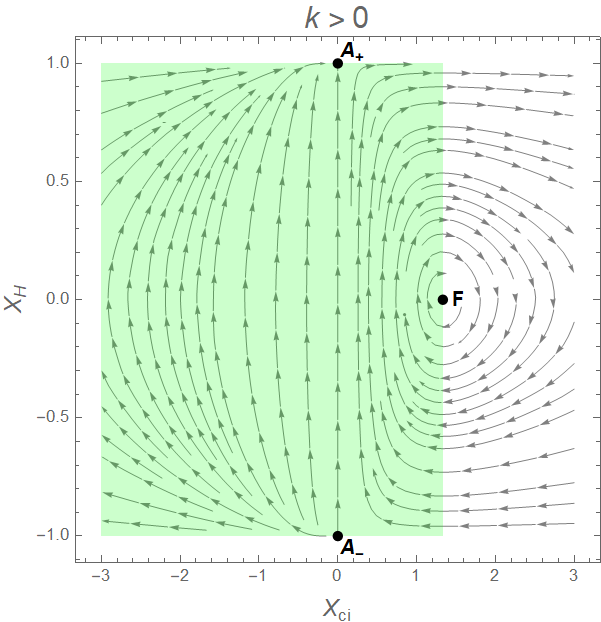}
    \includegraphics[width=0.294\textwidth]{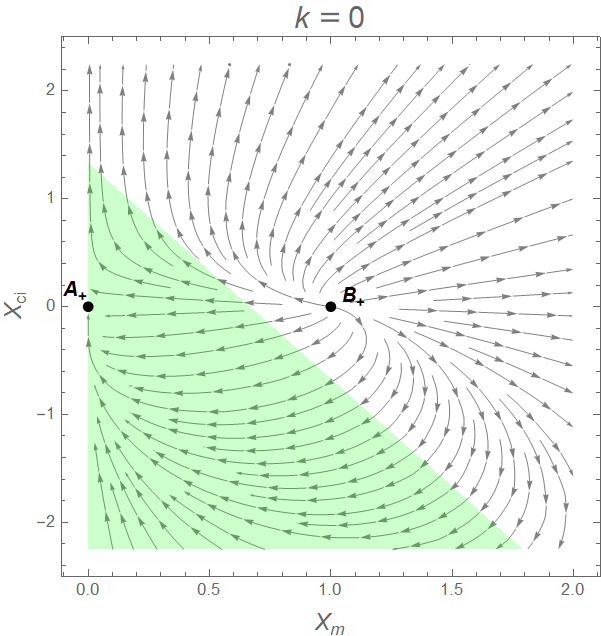}
  \caption{Invariant subset dominated by dark energy and spatial curvature (left and middle panels) and by matter and dark energy in the expanding case ($X_{H}=1$, right panel), with $w_{\rm ci0}=-0.5$ and $X_{\rm r}=0$. Critical points are indicated with black dots and the green shaded areas indicate accelerated phases of the dynamics ($q<0$).
 }
  \label{fig:XhXde0}
\end{figure*} 
    \item Set F: The only set of critical points of the system for positive spatial curvature (closed space) represents static solutions. The effective EoS parameter of the model is $w_{\rm eff}=-1/3$, while the scale factor of the static universe, obtained from \eqref{eq:friedpos}, is given by
\begin{equation}
 a = \exp{\left( \frac{1}{3(1+w_{\rm ci0})}-\frac{1-X_{\rm m}-X_{\rm r}}{2-3\, X_{\rm m}-4\, X_{\rm r}} \right)}.
 \label{eq:scaleF}
\end{equation}
If we consider only simple-gDE as a source of the static universe, one has to have $X_{\rm ci}=\frac{2}{3(1+w_{\rm ci0})}$ and hence $w_{\rm ci0}>-1$ if $X_{\rm ci}>0$. Under the same assumption of positive dark energy density, the set behaves as a center. In Fig.~\ref{fig:3d_perio}, we show the presence of periodic orbits evolving around the critical line $F$ for $X_{\rm r}=0$. These represent cyclic cosmological dynamics, periodically passing through phases of expansion and contraction without any past or future singularity.
\end{enumerate}

As already stated above, the stability of the de Sitter points $A_\pm$ depends on whether $w_{\rm ci0}$ is less or greater than $-1$. The same holds true for the existence of $E$ and $F$: assuming $X_{\rm ci}>0$, the former exists only for $w_{\rm ci0}<-1$, while the latter exists for $w_{\rm ci0}>-1$. Finally, the locations of $E$ and $F$ in the parameter space depend on the specific value of $w_{\rm ci0}$.

In Fig.~\ref{fig:XhXde0}, we show the trajectories of the system inside the invariant subsets $(X_{\rm ci},X_{H})$ and $(X_{\rm ci},X_{\rm m})$ for the case $w_{\rm ci0}>-1$: the corresponding case $w_{\rm ci0}<-1$ can be obtained by simply mirroring the plots through $X_{\rm ci}=0$.  Moreover, the value $w_{\rm ci0}=-1/2$ that we choose in order to plot the two-dimensional invariant subsets is only a representative value that allows a better visualization, but it is worth stressing that the topology of the trajectories and the stability character of the critical points do not change as long as $w_{\rm ci0}$ does not cross the value $-1$.

\subsection{Critical points at infinity}
The parameter space spanned by our dimensionless variables $X_i$ is not compact, and hence some trajectories can escape to infinity.  For instance, solutions in the rightmost panel of Fig.~\ref{fig:XhXde0}, emerging from the source $B_+$ and passing close to the unstable point $A_+$ in the upper half plane ($X_{\rm ci}>0$) will evolve toward $X_{\rm ci}\rightarrow\infty$.  This should represent phases of the cosmic evolution in which the dark energy contribution dominates over the other components, and hence it is important to understand the behavior of the system at infinity.  We consider the spatially flat expanding case, $X_{H}=1$, and we compactify the parameter space by defining the new variables $Y_i$ as
\begin{align}
 X_{\rm m} &= \frac{Y_{\rm m}}{Z}\quad ,\quad X_{\rm r} = \frac{Y_{\rm r}}{Z}\quad ,\quad X_{\rm ci} = \frac{Y_{\rm ci}}{Z} 
 \label{eq:compvar}
\end{align}
with $Y_{\rm m}^2+Y_{\rm r}^2+Y_{\rm ci}^2+Z^2=1$. 
The system at infinity corresponds to the unit sphere in the limit $Z\rightarrow0$, where the original variables blow up.  We recast the system of equations \eqref{eq:Xmp}--\eqref{eq:Xde0p} in terms of the variables \eqref{eq:compvar}, implement the constraint $Y_{\rm m}^2+Y_{\rm r}^2+Y_{\rm ci}^2+Z^2=1$ in order to get rid of $Y_{\rm ci}$, and take the limit $Z\rightarrow0$; the resulting reduced system at infinity is
\begin{equation}
\begin{aligned}
 Y_{\rm m}' &= Y_{\rm m} \left( 3 Y_{\rm m}^2 +4Y_{\rm r}^2 -3 \right),\\
 Y_{\rm r}' &= Y_{\rm r} \left( 3 Y_{\rm m}^2 +4Y_{\rm r}^2 -4 \right).
\end{aligned}
\end{equation}
By solving the system $Y_{\rm m}'=Y_{\rm r}'=0$ we find the following critical points:
\begin{itemize}
 \item $\left\{ Y_{\rm m}=1 , Y_{\rm r}=0 , Y_{\rm ci}=0 \right\}$
 \item $\left\{ Y_{\rm m}=0 , Y_{\rm r}=1 , Y_{\rm ci}=0 \right\}$
 \item $\left\{ Y_{\rm m}=0 , Y_{\rm r}=0 , Y_{\rm ci}=1 \right\}$.
\end{itemize}
\begin{figure*}[ht!]
 \includegraphics[width=0.24\textwidth]{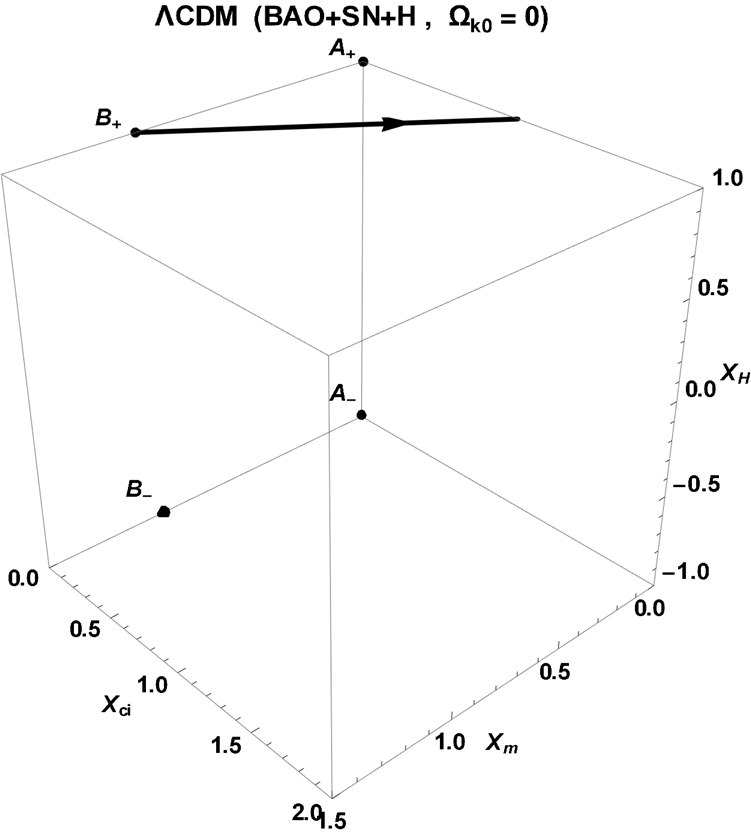}%\hspace{0.4cm}
 \includegraphics[width=0.24\textwidth]{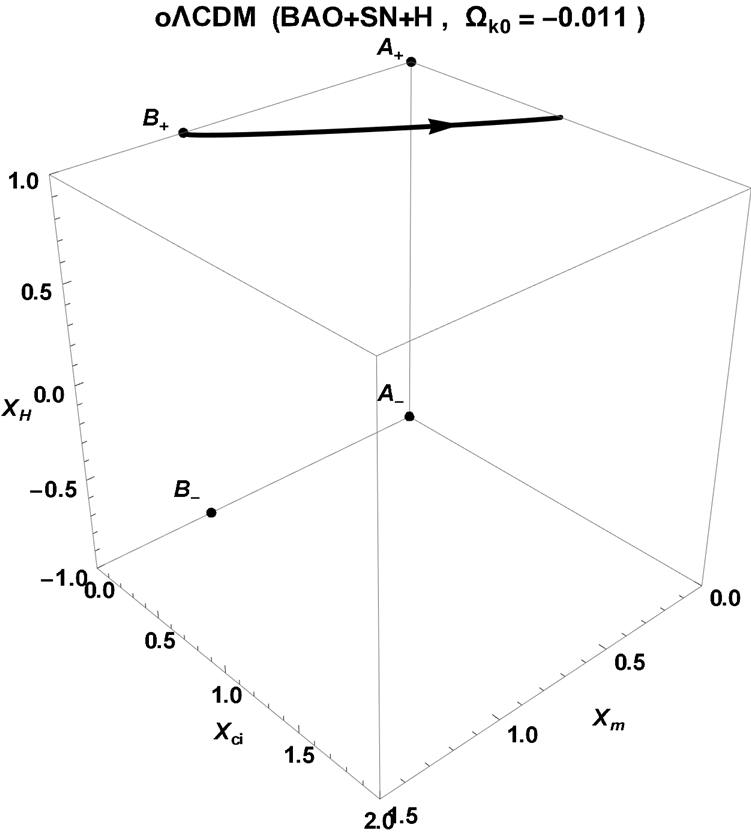}%\hspace{0.4cm}
 \includegraphics[width=0.24\textwidth]{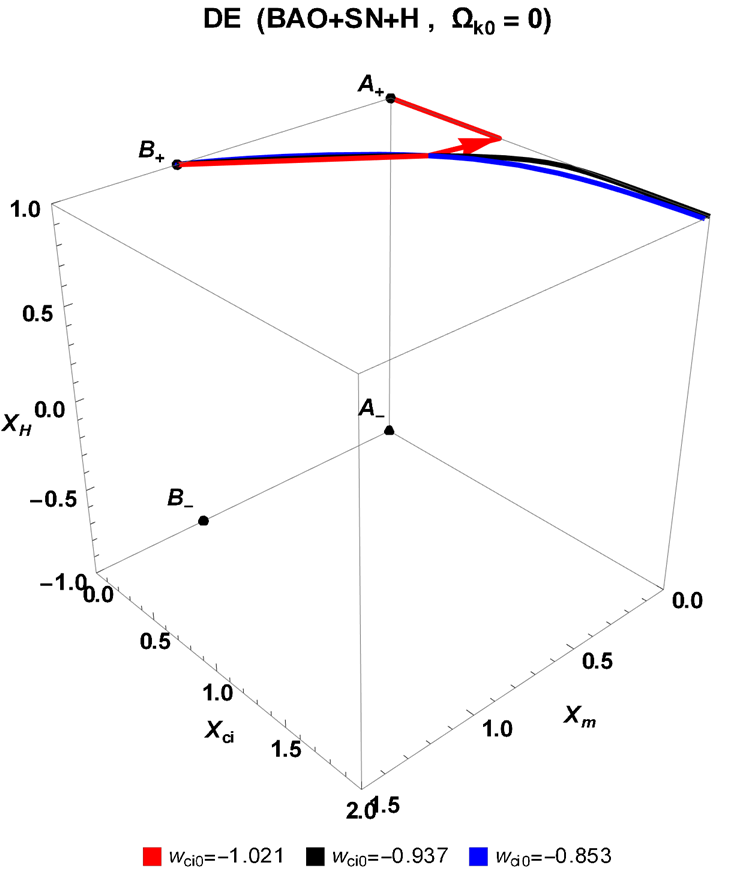}%\hspace{0.4cm}
 \includegraphics[width=0.24\textwidth]{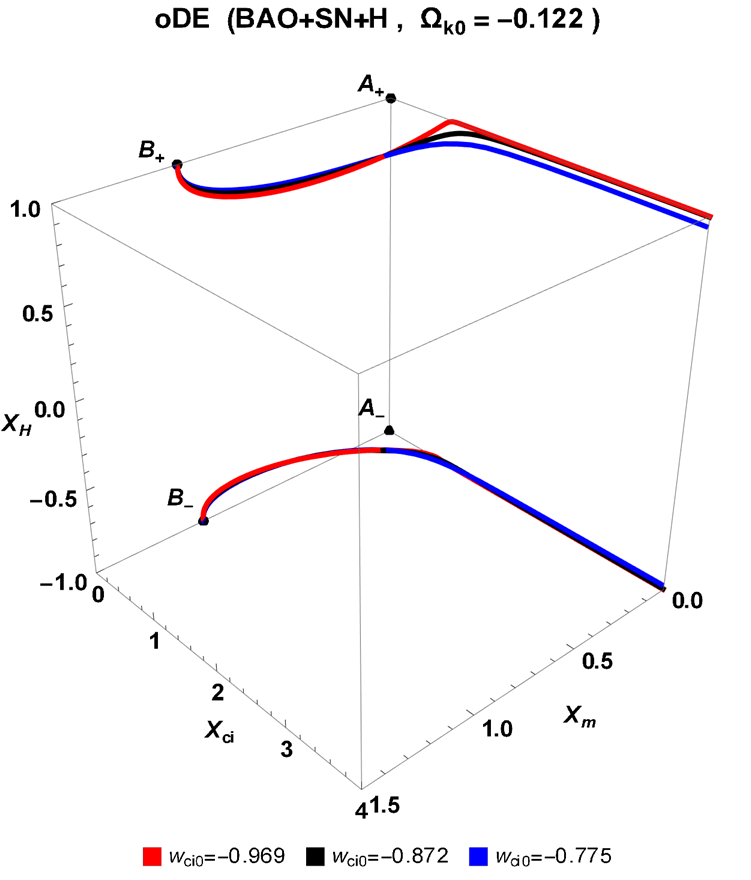}
  \caption{Trajectories corresponding to the $\Lambda$CDM, $o\Lambda$CDM, DE, and $o$DE models, for values of the parameter $w_{\rm ci0}$ in the allowed range according to the combined BAO+SN+$H$ dataset given in Table~\ref{tab:priors}.}
  \label{fig:3d_tab1_noPLK}
\end{figure*}

\begin{figure*}[ht!]
 \includegraphics[width=0.23\textwidth]{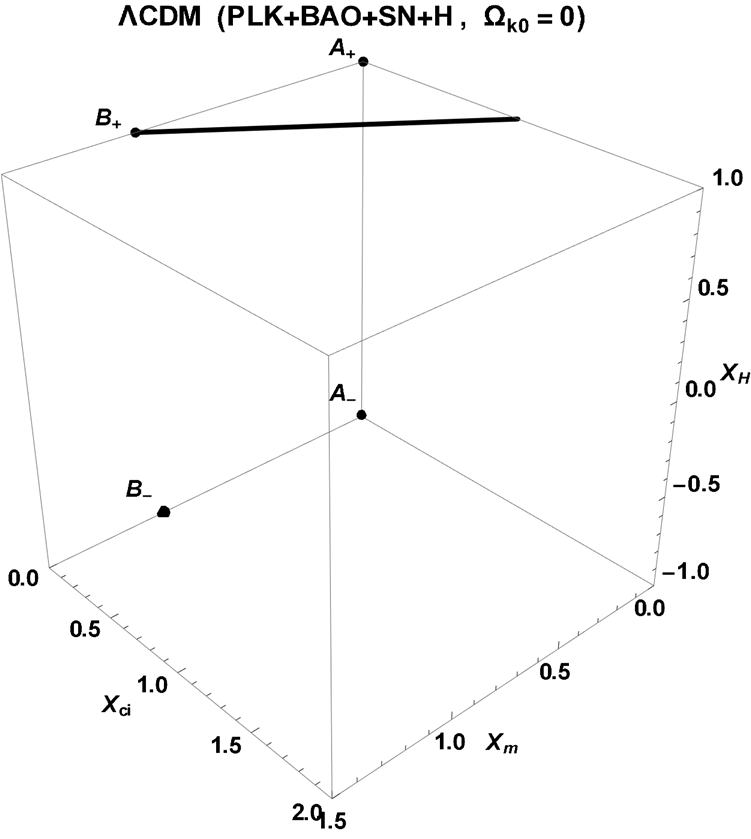}%\hspace{0.5cm}
 \includegraphics[width=0.23\textwidth]{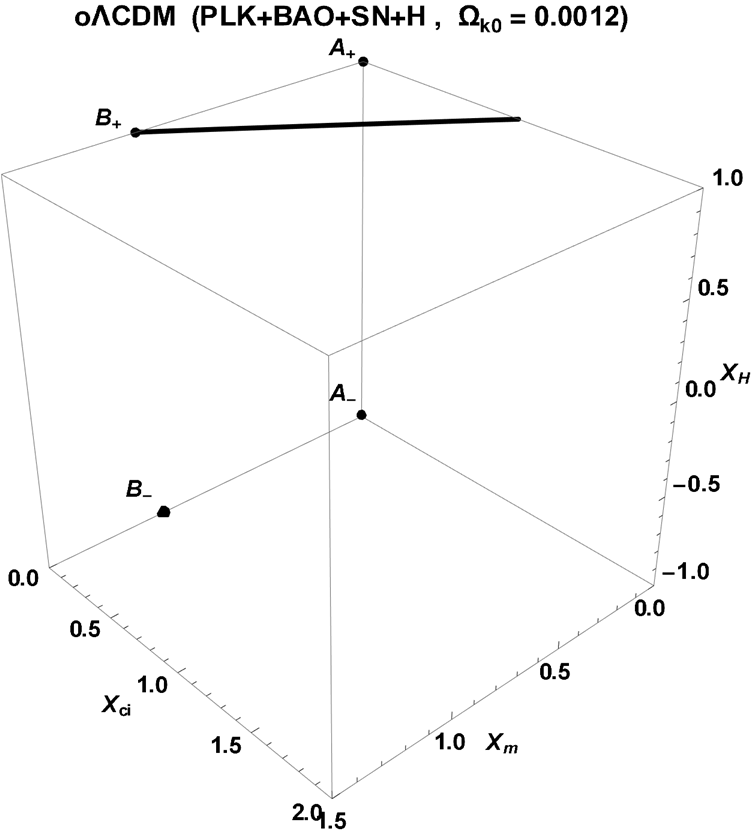}
 \includegraphics[width=0.23\textwidth]{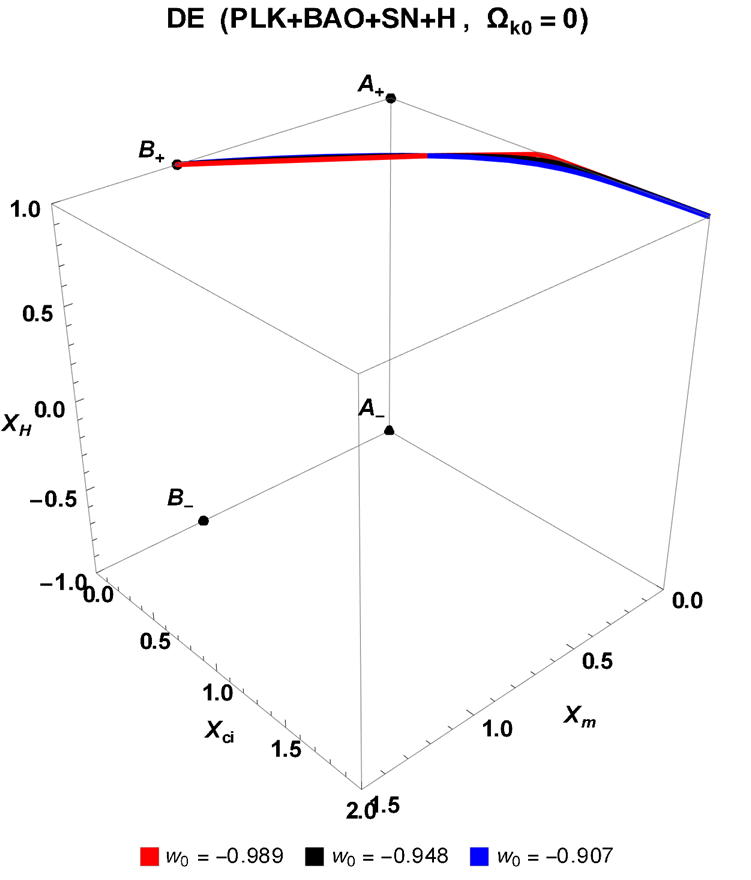}%\hspace{0.5cm}
 \includegraphics[width=0.23\textwidth]{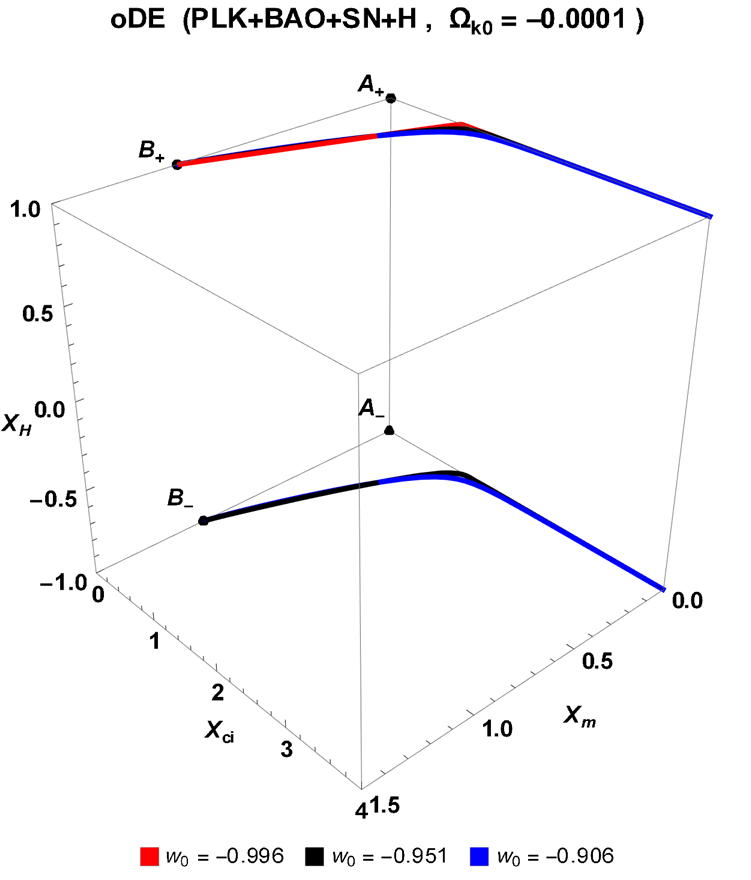}
  \caption{Trajectories corresponding to the $\Lambda$CDM, $o\Lambda$CDM, DE, and $o$DE models, for values of the parameter $w_{\rm ci0}$ in the allowed range according to the combined BAO+SN+$H$+PLK dataset given in Table~\ref{tab:priors}.}
  \label{fig:3d_tab1_PLK}
\end{figure*}
The last point is of particular interest, because $Y_{\rm ci}$ dominates over the other variables.  In terms of the old variables, this means that $X_{\rm ci}$ dominates over $X_{\rm m}$ and $X_{\rm r}$, and hence the Friedmann equation takes the form
\begin{equation}
\begin{aligned}
 1 &=X_{\rm ci} \left[ 1-3\, (1+w_{\rm ci0})\, \ln a \right]\\
 &=\frac{\rho_{\rm ci0}}{3} \frac{a^2}{\dot{a}^2} \left[ 1-3\, (1+w_{\rm ci0})\, \ln a \right]\, .
\end{aligned}
\end{equation}
Solving this equation for $a(t)$ with initial condition $a(t_0)=1$, one gets
\begin{equation}\label{eq:scale_DE}
 a = \exp \left[ \sqrt{\frac{\rho_{\rm ci0}}{3}}\, (t-t_0)-(1+w_{\rm ci0})\frac{\rho_{\rm ci0}}{4}\, (t-t_0)^2 \right]\, .
\end{equation}
For $w_{\rm ci0}>-1$, this model describes a bounce, with maximum scale factor $a={\rm e}^{-\frac{1}{3(1+w_{\rm ci0})}}$; for $w_{\rm ci0}<-1$ it describes the LSBR, which is an abrupt event first proposed in Ref. \cite{Bouhmadi-Lopez:2014cca}, for which the Hubble parameter diverges, whereas its time derivative remains finite:
\begin{align}
 H &= \sqrt{\frac{\rho_{\rm ci0}}{3}}-\frac{(1+w_{\rm ci0})\, \rho_{\rm ci0}}{2}\, (t-t_0)\, ,\\
 \dot{H} &=-\frac{(1+w_{\rm ci0})}{2}\, \rho_{\rm ci0}=-\frac{1}{2}\varrho_{\rm ci}\, .
\end{align}
In the spatially flat expanding case and considering a dark-energy-dominated phase, the system reduces to
\begin{equation}
 X_{\rm ci}' = 3\, (1+w_{\rm ci0})\, X_{\rm ci}^2\, .
\end{equation}
Hence, for $w_{\rm ci0}>-1$ ($\varrho_{\rm ci}>0$), the point at infinity is a future attractor bouncing scenario, while for $w_{\rm ci0}<-1$ ($\varrho_{\rm ci}<0$), it is a past attractor LSBR model.  In the next section we will show numerically that, although the presence of curvature renders such point at infinity an unstable saddle, the dynamics for $w_{\rm ci0}>-1$ ($\varrho_{\rm ci}>0$) still leads to a recollapsing scenario.

\subsection{Dynamics constrained by observations}

At this point, we are ready to implement into the dynamical system the values of the cosmological parameters offered by the datasets in Table~\ref{tab:priors} in order to identify the trajectories and late-time attractors of the system under observational constraints.  As the initial conditions for the evolution of our dynamical system, we consider the mean values as well as their deviations within the 1$\sigma$ error bar, by relating the fractional densities to the $X_i$ variables according to the following:
\begin{align}
\Omega_{\rm m}=&\frac{X_{\rm m}}{X_{H}^2}\quad ,\quad \Omega_{\rm r} = \frac{X_{\rm r}}{X_{H}^2}\quad,\quad \Omega_{\rm ci} = \frac{X_{\rm ci}}{X_{H}^2},
\end{align}
\begin{align}
\Omega_k =&\left\{
  \begin{array}{@{}ll@{}}
    -\frac{1}{3}\left(1-\frac{1}{X_{H}^2}\right)\quad &\text{for}\ k\leq0,\\
    \frac{1}{3}\left(1-\frac{1}{X_{H}^2}\right)\quad &\text{for}\ k>0.
  \end{array}\right. 
  \label{eq:omegak}
\end{align}
The current state of the universe being in an expanding phase, we choose $X_{H}>0$ at present day.  As a consequence, we can expect the trajectory to be in the basin of past attraction of the radiation-dominated point $C_+$; the trajectory emerges from there and passes close to the matter-dominated saddle $B_+$, the second epoch of the cosmological evolution. 

Assuming $\rho_{\rm ci0}>0$, we see that (i) for $w_{\rm ci0}>-1$ ($\varrho_{\rm ci}>0$) the recollapsing scenario takes place irrespective of spatial curvature and (ii) for $w_{\rm ci0}<-1$ ($\varrho_{\rm ci}<0$) the asymptotic future attractor is point $A_+$, the de Sitter Universe.
 
In Figs.~\ref{fig:3d_tab1_noPLK} and \ref{fig:3d_tab1_PLK}, we plot the trajectories corresponding to the cosmological parameters constrained, respectively, by the combined datasets of BAO+SN+$H$ and BAO+SN+$H$+PLK, given in Table~\ref{tab:priors}. The first two panels of these figures show the cosmic evolution for $w_{\rm ci0}=-1$, for the $\Lambda$CDM and $o\Lambda$CDM models; the asymptotic future behavior is the usual cosmological constant-dominated phase with $X_{\rm ci}=1$.  In the remaining panels of the same figures, we plot the trajectories for the DE model and $o$DE model, highlighting three trajectories corresponding to the mean value of $w_{\rm ci0}$ and its upper and lower values in the error bar region.  From the values provided by Table~\ref{tab:priors}, it is clear that both datasets (with or without Planck CMB data) favor $w_{\rm ci0}>-1$  ($\varrho_{\rm ci}>0$), and hence the recollapsing model is a generic future behavior; in the DE model, it appears as a future attractor at infinity, while when spatial curvature is allowed in the $o$DE model the recollapse takes place due to the change in sign of $X_{H}$. For $w_{\rm ci0}<-1$  ($\varrho_{\rm ci}<0$) (not favored, but is still a possibility, by the observational data), on the other hand, the de Sitter point $A_+$ is a future attractor; the DE model constrained without Planck CMB data is the only one that allows such a scenario within 1$\sigma$ confidence region (see the third panel of Fig.~\ref{fig:3d_tab1_noPLK}).

\section{Conclusions}
\label{sec:conclusions}
In this paper, we first discussed briefly the possibility that dark energy models with negative energy density values in the past can alleviate the $H_0$ tension, as well as the discrepancy with the Ly-$\alpha$ BAO measurements, both of which prevail within the standard $\Lambda$CDM model. We have then investigated in detail whether two minimal extensions of the $\Lambda$CDM model, together or separately, can successfully realize such a scenario: (i) the spatial curvature, which, in the case of spatially closed universe, mimics a negative energy density source with an EoS parameter $w=-1/3$ and (ii) simple-graduated dark energy, which promotes the null inertial mass density of the usual vacuum energy to an arbitrary constant as $\varrho=\rm const$ (i.e., the minimal deviation from null value), which, if negative, results in the corresponding energy density decreasing with increasing redshift similar to the phantom models, but unlike them crossing below zero at a certain redshift---see the \textit{graduated dark energy} model \cite{Akarsu:2019hmw}  for the minimal dynamical deviation from the null inertial mass density. We have found that, when the Planck CMB data are not included in the observational analysis using the combined BAO+SN+$H$ dataset, the models with simple-gDE predict interesting and some significant deviations from the $\Lambda$CDM model. In particular, this dataset predicts $\Omega_{k0}$ and $\varrho_{\rm ci}$ that oppose each other in enhancing $H_0$, namely, predicts a spatially closed universe of $\Omega_{k0}=-0.122\pm0.117$ along with a simple-gDE of a positive inertial mass density $\varrho_{\rm ci}= (7.65\pm5.72)\times 10^{-31} \, {\rm g\, cm^{-3}}$, which results in $H_0=68.84\pm2.60\,{\rm km\,s}^{-1}{\rm Mpc}^{-1}$ implying no robust improvement in the $H_0$ tension---notice the increased errors of this value with respect to the ones, $H_0=68.27\pm0.88\,{\rm km\,s}^{-1}{\rm Mpc}^{-1}$, in the case of the $\Lambda$CDM model. Besides, the Bayesian evidence suggests that there is a significant evidence for preferring the $\Lambda$CDM model over the extended models. The combined BAO+SN+$H$+PLK dataset, including the Planck CMB data, presents significant evidence against the $o\Lambda$CDM and $o$DE models, i.e., a deviation from spatial flatness in both models considering the $\Lambda$ and the simple-gDE as the dark energy sources, but presents almost the same evidence for the $\Lambda$CDM model and the DE model (simple-gDE) with a positive inertial mass density at the order of $\mathcal{O}(10^{-12})\,\rm eV^4$, namely, $\varrho_{\rm ci}= (3.06\pm2.28)\times 10^{-31} \, {\rm g\, cm^{-3}}$. It is striking that this constraint implies statistically a significant deviation from the usual vacuum energy ($\varrho=0$), but, alas, with a sign opposite to our original expectation that would help alleviate the $H_0$ tension. This, however, predicts almost no deviation from the $\Lambda$CDM model up until today---so that it does not result in any improvement regarding the Ly-$\alpha$ BAO measurements--- except that it predicts slightly lower constraint $H_0=67.73\pm0.99\,{\rm km\,s}^{-1}{\rm Mpc}^{-1}$ on the Hubble constant compared to one $H_0=68.29\pm0.52\,{\rm km\,s}^{-1}{\rm Mpc}^{-1}$ obtained for the $\Lambda$CDM model and thereby aggravates the $H_0$ tension. We have also studied via dynamical analysis the complete history of the Universe in the models, in particular, as the simple-gDE results in two distinct futures depending on the sign of the inertial mass density different than the de Sitter future of the $\Lambda$CDM model. It turns out that the recollapse of the Universe in the finite future is a generic behavior of the simple-gDE models, as $\varrho>0$ within 68\% CL independent of whether the Planck data are included or not in the observational analyses---with the exception of the spatially flat simple-gDE case (DE) constrained without Planck CMB data, which allows a small incursion of $\varrho$ into the negative value region. Finally, it is surprising to observe, via simple-gDE---which promotes phenomenologically the null inertial mass density of the usual vacuum energy of QFT to an arbitrary constant---that the data favor a positive constant inertial mass density ($\varrho={\rm const}>0$), rather than a constant positive energy density ($\Lambda>0$), as one of the constants of nature subject to measurements. Therefore, we find it tempting to investigate whether such a generalization of the usual vacuum energy can be derived/predicted from a fundamental theory of physics.

\begin{acknowledgments}
This work is dedicated to the memory of Professor John David Barrow. \"{O}.A. acknowledges the support by the Turkish Academy of Sciences in scheme of the Outstanding Young Scientist Award  (T\"{U}BA-GEB\.{I}P). N.K. acknowledges the COST Action CA15117 (CANTATA). J.A.V. acknowledges the support provided by FOSEC SEP-CONACYT Investigaci\'on B\'asica A1-S-21925, FORDECYT-PRONACES-CONACYT/304001/2020 and UNAM-DGAPA-PAPIIT IA104221.
\end{acknowledgments}

\appendix
\bigskip
\bigskip
\end{document}